\documentclass[sigconf, nonacm]{acmart}
\usepackage[normalem]{ulem}
\usepackage{color}
\usepackage{listings}
\usepackage[]{hyperref}
\usepackage{amsmath,amsfonts}
\usepackage{mathtools}
\usepackage[ruled, vlined, linesnumbered]{algorithm2e}
\usepackage{circledsteps}
\usepackage{comment}
\usepackage{booktabs}
\usepackage{stmaryrd}
\usepackage{wasysym}
\usepackage{multirow}
\usepackage{graphicx}
\usepackage{makecell}
\usepackage{subcaption}
\usepackage{xcolor}
\usepackage{mathpartir}
\usepackage{adjustbox}
\usepackage{mdframed}
\usepackage{siunitx}
\usepackage{enumitem}
\usepackage{xcolor}
\usepackage{xstring}
\usepackage{marginnote}

\AtBeginDocument{%
  }

\newcommand{\Name}{ONEX\xspace}

\begin{document}

\title{Architecture and Compilation Co-Design for High-Rate Quantum Product Codes on Neutral Atom Arrays}

\author{Adrian Liu}
\email{adrianliu@ucla.edu}
\affiliation{
  \institution{University of California, Los Angeles}
  \city{Los Angeles}
  \state{CA}
  \country{USA}
}

\author{Wan-Hsuan Lin}
\email{wlin@quera.com}
\affiliation{
  \institution{QuEra Computing Inc.}
  \city{Boston}
  \state{MA}
  \country{USA}
}

\author{Daniel Bochen Tan}
\email{danieltan@g.harvard.edu}
\affiliation{
  \institution{Harvard University}
  \city{Cambridge}
  \state{MA}
  \country{USA}
}

\author{Qian Xu}
\email{qianxu@caltech.edu}
\affiliation{
  \institution{California Institute of Technology}
  \city{Pasadena}
  \state{CA}
  \country{USA}
}

\author{Jason Cong}
\email{cong@cs.ucla.edu}
\affiliation{
  \institution{University of California, Los Angeles}
  \city{Los Angeles}
  \state{CA}
  \country{USA}
}


%%%%%% -- PAPER CONTENT STARTS-- %%%%%%%%

\begin{abstract}
Achieving fault-tolerant quantum computing at a practical scale demands quantum error correction (QEC) codes with high encoding rates. Quantum low-density parity-check (qLDPC) codes emerge as a promising candidate, especially given the rise of neutral atom arrays that provide dynamic long-range connectivity via atom movements. In general, synthesizing valid and efficient physical execution plans for QEC is a provably hard combinatorial problem, forming a critical compilation bottleneck that worsens as code sizes grow. 
To overcome this complexity, we focus on an important product family of qLDPC codes with dimension-reduction properties, and propose \Name. This framework decomposes complex 2D physical execution planning into independent 1D subproblems, each solved to optimal execution depth within practical compilation time.
% In particular, \Name introduces four key contributions. 
First, we formulate the 1D execution plan with an explicit satisfiability modulo theories (SMT) encoding. This protocol produces provably depth-optimal solutions with substantial duration reduction.
Second, we develop a multi-stage compilation pipeline featuring anytime optimization, movement compaction, and iterative feedback. This pipeline maintains practical wall-clock times while providing progressive refinement and on-demand retrieval of quality solutions.
Third, we evaluate \Name in the application of hypergraph product (HGP) code memory mapped onto neutral atom arrays, achieving 3.7$\times$ to 6.1$\times$ and 29.8$\times$ to 42.1$\times$ higher clock rates than the constructive 1D algorithm and the general 2D compiler, respectively, while scaling efficiently to codes with 2{,}500 data qubits. 
Finally, we extend \Name to zoned layouts, revealing architectural insights into the associated trade-offs, and demonstrate its applicability to the broader lifted-product (LP) code family through a representative example.

% \Name is open-source and publicly available.

\end{abstract}
%\begin{CCSXML}
%<ccs2012>
%  <concept_id>00000000.0000000.0000000</concept_id>
%  <concept_desc>Do Not Use This Code, Generate the Correct Terms for Your Paper</concept_desc>
%  <concept_significance>500</concept_significance>
%  %<concept_id>00000000.00000000.00000000</concept_id>
%  <concept_desc>Do Not Use This Code, Generate the Correct Terms for Your Paper</concept_desc>
%  <concept_significance>300</concept_significance>
%  %<concept_id>00000000.00000000.00000000</concept_id>
%  <concept_desc>Do Not Use This Code, Generate the Correct Terms for Your Paper</concept_desc>
%  <concept_significance>100</concept_significance>
 % <concept_id>00000000.00000000.00000000</concept_id>
%  <concept_desc>Do Not Use This Code, Generate the Correct Terms for Your Paper</concept_desc>
%  <concept_significance>100</concept_significance>
%</ccs2012>
%\end{CCSXML}

%\ccsdesc[500]{Do Not Use This Code~Generate the Correct Terms for Your Paper}
%\ccsdesc[300]{Do Not Use This Code~Generate the Correct Terms for Your Paper}
%\ccsdesc{Do Not Use This Code~Generate the Correct Terms for Your Paper}
%\ccsdesc[100]{Do Not Use This Code~Generate the Correct Terms for Your Paper}


\maketitle

\section{Introduction}
% \cmt{The Need: High-Rate Quantum Error Correction}

% \hyphenation{implies}
% \hyphenation{practical}
% \hyphenation{physical}

Fault-tolerant quantum computing (FTQC) demands quantum error correction (QEC) to protect logical information from physical noise~\cite{faulttolerantwithconstanterror,bluvstein_logical_2024,Nielsen_Chuang_2010,acharya_suppressing_2023,sivak_real-time_2023}.
The well-known surface code~\cite{surfacecode} achieves a high error threshold but yields an encoding rate that vanishes as $\mathcal{O}(1/d^2)$ with code distance $d$. This poor encoding efficiency implies enormous resource overhead, suggesting that millions of physical qubits may be necessary to execute algorithms of practical interest~\cite{gidney_how_2021,gidney_how_2025}.
In contrast, high-rate quantum low-density parity-check (qLDPC) codes offer a fundamentally different scaling trajectory~\cite{hgp,hgpcode,lpcode,bbcode}.
% By combining constant encoding rates with large code distances, these qLDPC codes have the potential to reduce the physical overhead by orders of magnitude, making practical applications promising with around 10{,}000 physical qubits~\cite{cain2026possible10000}.
By combining constant encoding rates with large code distances, these qLDPC codes have the potential to reduce the physical overhead by orders of magnitude, thereby making practical applications promising even with around 10{,}000 physical qubits~\cite{cain2026possible10000}.
As such, high-rate qLDPC codes represent a viable path toward resource-efficient QEC at a practical scale and arise as a major focus of recent research.

\begin{figure*}[t]
\centering
\includegraphics[width=0.9\linewidth]{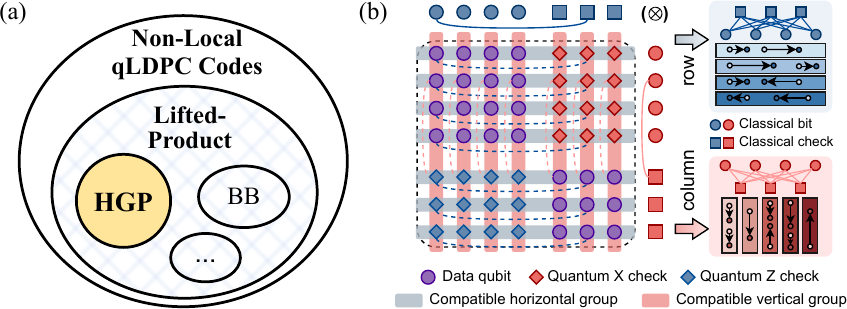}
\caption{
% Motivation. (a)~Product construction underpins many high-rate qLDPC families, with HGP codes as a primary showcase. (b)~The Cartesian decomposition of HGP syndrome extraction into independent 1D subproblems, allowing for parallel execution of compatible groups.
Motivation. (a)~Product construction underpins many high-rate qLDPC families, with HGP codes as a primary showcase.
(b)~HGP syndrome extraction on a product-aligned neutral-atom layout. Purple circles are data-qubit atoms; red/blue diamonds are $X/Z$ check-ancilla atoms. Each classical seed code is replicated along one product coordinate, yielding horizontal groups (gray) and vertical groups (pink). The right panels show the corresponding 1D row/column subproblems; small bars denote a 1D execution plan with rearrangement steps. Compatible groups execute in parallel.
}
\label{fig:intro}
\Description{}
\end{figure*}

Distinct from geometrically local topological codes, prominent qLDPC codes require non-local qubit interactions beyond nearest-neighbor connectivity to implement syndrome extraction.
Neutral atom arrays have emerged as one of the most promising platforms for meeting this requirement. By trapping individual atoms in reconfigurable optical tweezer arrays, these systems provide dynamic, long-range connectivity through physical atom rearrangement \cite{bluvstein_quantum_2022,saffman_quantum_2018,Henriet2020quantumcomputing}. 
Recent experiments have demonstrated processors with several thousand qubits \cite{Pause24,cavityenhanced,manetsch_tweezer_2025}, positioning this platform at the forefront of large-scale quantum computation. 

% \cmt{The Challenge: Optimal Compilation}

% Cosmetic first-page adjustment; revisit if preceding text changes.
However, the flexibility of atom rearrangement introduces a formidable compilation challenge to achieving efficient execution of QEC, particularly the syndrome extraction.
Physical constraints of acousto-optic deflector (AOD) control render rearrangement synthesis a provably NP-hard combinatorial problem even in one dimension \cite{laguna_grasp_1999,junger_2-layer_1997}. To manage such complexity, existing general compilation workflows often compromise by partitioning the rearrangement problem into separate placement and routing stages that converge prematurely to local minima \cite{enola,zac,azac}. 
For many qLDPC instances, even with near-optimal scheduling that minimizes circuit depth~\cite{enola}, general-purpose compilations can still produce solutions more than 40$\times$ slower than the highly optimized solutions from this work (see Section~\ref{sec:hgp_app}).
As the dominant component of execution duration, suboptimal rearrangement directly limits the clock rate of the quantum processor. 
Other studies design code-specific layouts that leverage particular symmetries to achieve efficient rearrangement under restricted patterns \cite{qsieve,Viszlai_2025,Coniq,yang_rascql_2026,wang_coprime_2026} but do not generalize to instances outside these assumptions, including those studied here, yet a vast design space of broader qLDPC code classes still remains as an open challenge to be explored.
% Other studies design code-specific layouts that leverage particular symmetries to achieve efficient rearrangement under restricted patterns \cite{qsieve,Viszlai_2025,Coniq,yang_rascql_2026,wang_coprime_2026} and do not generalize to instances such as those studied here, yet a vast design space of broader qLDPC code classes still remains as an open challenge to be explored.

% Other studies design code-specific layouts that exploit particular symmetries for efficient rearrangement under restricted patterns \cite{qsieve,Viszlai_2025,Coniq,yang_rascql_2026,wang_coprime_2026} but do not generalize to instances such as those studied here, leaving broader qLDPC code classes largely unexplored.

% \cmt{Structural Opportunity}

Within this landscape, a notable structural commonality among many prominent qLDPC code families is their construction from generalized products of classical component codes \cite{hgpcode, lpcode, bbcode} (Fig.~\ref{fig:intro}a).
Critically, the product construction induces a natural decomposition of stabilizer interactions into independent groups aligned along orthogonal dimensions. This decomposition has direct architectural implications, as it determines how two-qubit gates can be organized and parallelized during syndrome extraction. 
Nevertheless, no existing framework fully leverages these structural opportunities for holistic physical execution synthesis, which requires co-optimizing qubit mapping and gate execution alongside atom rearrangement.
% Current heuristic methods focus primarily on rearrangement scheduling and remain limited in the quality of solutions they produce \cite{xu_constant-overhead_2024}.
% Current heuristics focus narrowly on local rearrangement generation without global consideration and remain limited in the quality of solutions they produce \cite{xu_constant-overhead_2024}. Experimental results show that the asymptotically optimal constructive heuristic can still produce solutions more than 6$\times$ worse than the true optimum (see Section~\ref{sec:hgp_app}).
Existing heuristics generate rearrangements locally without global coordination, limiting solution quality; even under the same structural decomposition, the asymptotically optimal local heuristic~\cite{xu_constant-overhead_2024}, can still require up to 9$\times$ as many rearrangement steps in practice as the exact depth optimum obtained in this work (see Section~\ref{sec:1d_app}).

% produce solutions with 7$\times$ more execution depth than the true optimum (see Section~\ref{sec:hgp_app}).}

% Existing heuristic-based method, though valid, only focus on the rearrangement schedule and remains limited in the quality of solutions they produce \cite{xu_constant-overhead_2024}.

% In this work, we focus on HGP codes as our primary showcase, since their structure is purely product-based, while other product-code families carry analogous factorization properties alongside additional algebraic structure.

% \cmt{Our Solution}

In this paper, we propose \Name (\underline{O}ptimal dimensional \underline{N}eutral-atom \underline{Ex}ecution compiler), an architecture and compilation co-design framework for scalable high-rate quantum error correction on neutral atom arrays.
\Name exploits the structure of product qLDPC code families, by naturally decomposing stabilizer interactions into two independent groups that map directly onto the Cartesian control axes of AODs, as illustrated with hypergraph product (HGP) codes~\cite{hgpcode} in Fig.~\ref{fig:intro}b.
% \Name exploits the product structure of qLDPC codes, such as hypergraph product (HGP) codes \cite{hgpcode}, by naturally decomposing stabilizer interactions into two independent groups that map directly onto the Cartesian control axes of AODs (Fig.~\ref{fig:intro}b).
Our key insight is that this correspondence allows the complex 2D physical execution planning to be decomposed into orthogonal parallel 1D subproblems that are solvable optimally, enabling us to reduce the compilation complexity from $\mathcal{O}(2^{n^2})$ to $\mathcal{O}(2^{n})$. This approach also aligns with recent practical efforts to treat structural decomposability as a design criterion for ultra-high-rate codes \cite{ultra_high_rate}.

Our focus on qLDPC codes with product structure is analogous to the choice of slicing floorplans in VLSI layout design in the 1980s. Despite being a subset of general floorplans, slicing floorplans offered an efficient encoding and a compact search space~\cite{stockmeyer_optimal_1984}, which often produced solutions superior to those from general floorplans and became widely adopted in practice~\cite{wong_new_1986,young_slicing_1999}.

Particularly, this paper makes the following contributions:
\begin{itemize}[leftmargin=*]
    \item \textbf{A highly efficient 1D execution protocol.} We formulate the optimal 1D atom execution planning into  formal Satisfiability Modulo Theories (SMT) abstraction. This protocol produces depth-optimal solutions with an 8.4$\times$ improvement in duration compared to the baseline \cite{xu_constant-overhead_2024}, enabling higher clock rates and suppressed error accumulation.
    
    \item \textbf{A multi-stage compilation framework.} We develop a three-phase pipeline that integrates anytime solving, movement compaction, and iterative feedback with multi-level parallelism. This framework ensures progressive solution refinement with on-demand retrieval and practical wall-clock compilation times.

    \item \textbf{Application-level QEC mapping and evaluation.} We demonstrate high-rate HGP memory architectures on neutral atom arrays, achieving clock rates 3.7$\times$ to 6.1$\times$ above the 1D constructive routing algorithm and 29.8$\times$ to 42.1$\times$ above the general 2D compiler, while scaling to codes exceeding 2{,}500 data qubits with broadly improved logical error rates. Generalization to the broader lifted-product (LP) code family is also studied and demonstrated on a recent advance~\cite{cain2026possible10000}.

    \item \textbf{Architectural adaptation to zoned layouts.} We extend \Name to zoned layouts and compare multiple execution strategies against existing zoned compilers. To the best of our knowledge, this is the first study to analyze the inter- and intra-zone trade-off for QEC execution on zoned neutral-atom processors.

\end{itemize}

% The rest of the paper is organized as follows. Section~\ref{sec:preli} provides the necessary background. Section~\ref{sec:overview} gives an overview of \Name and Section~\ref{sec:method} details our methodology. Section~\ref{sec:eval} presents evaluation results and Section~\ref{sec:discussion} discusses broader potential and current limitations. Finally, Section~\ref{sec:conclusion} concludes the paper.

\section{Preliminaries} \label{sec:preli}
This section introduces the neutral-atom hardware model and the structured quantum error correction codes that together motivate the co-design approach of \Name.

%% ---------------------------------------------------------------------------
\subsection{Neutral-Atom Quantum Computing}
\label{sec:preli-na}

\begin{figure}[t]
\centering
\includegraphics[width=\linewidth]{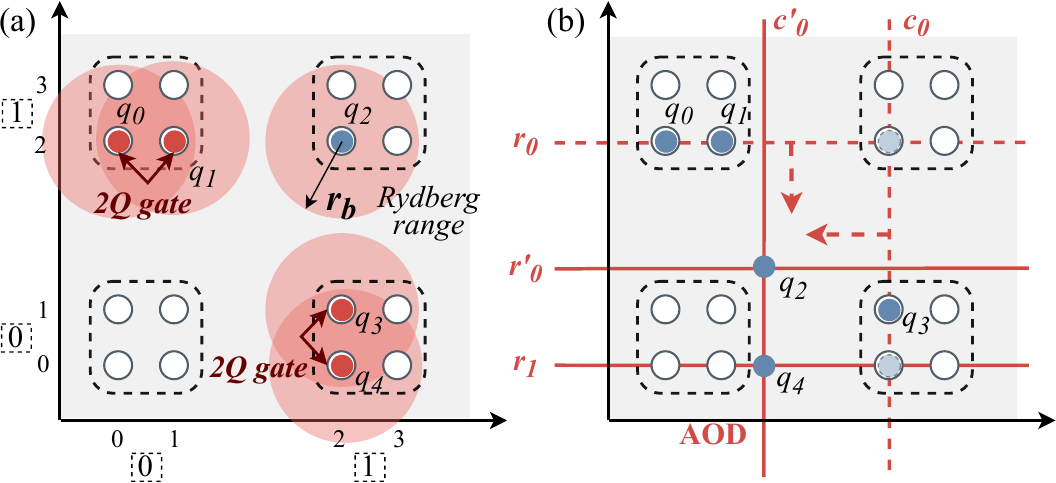}
\caption{Neutral-atom hardware model. (a)~A 2D optical tweezer array with atoms (dots) held at traps (circles). 
Adjacent trap pairs form interaction sites (dashed box) where Rydberg entangling gates execute. (b)~Intersectional atom transport via row- and column-axis AODs.}
\label{fig:hardware}
\Description{}
\end{figure}

Neutral-atom quantum processors trap individual atoms in tightly focused laser beams known as optical tweezers, arranged in a two-dimensional grid (Fig.~\ref{fig:hardware}a). Static tweezers generated by a spatial light modulator~(SLM) define fixed traps, while dynamic tweezers driven by AODs transport atoms between traps. 
Entangling gates are executed via Rydberg blockade by positioning two atoms within a shared \emph{interaction site}, where adjacent traps reside within each other's illumination zone to enable controlled operations \cite{rydberg}.
Based on the presence of dedicated storage zones, neutral-atom processors are categorized into \emph{monolithic} and \emph{zoned} architectures, each requiring specialized compilation efforts \cite{stade_logical_zone, zac, azac, powermove, stade2025search, olsq_dpqa, olsq_dpqa_2, atomique, enola, dasatom, tan_compiling_2024}. Here, \Name first focuses on the monolithic case and then discusses its adaptation to zoned layouts in Section~\ref{sec:zoned}.

The AOD control geometry is illustrated in Fig.~\ref{fig:hardware}b. Each AOD channel deflects atoms along a single axis (row or column), with mobile traps formed at the intersection of active row and column channels. 
All atoms sharing a channel move simultaneously and the coordination of multiple AODs imposes a strict \emph{no-crossing constraint}. Specifically, AOD channels in the same dimension (\textit{e.g.}, $r_0$ and $r_1$) cannot cross each other. This ordering-preservation rule prevents atoms from heating up. This requirement distinguishes neutral-atom rearrangement from general permutation and is the primary source of the problem's combinatorial complexity.

The rearrangement of atoms between gate stages incurs two primary costs. \emph{Execution depth} counts the number of discrete rearrangement steps, each consisting of a three-phase sequence:
\begin{enumerate}[leftmargin=*]
    \item \textit{activate}: atoms transfer from static SLM traps to the dynamic AOD control grid, enabling associated movement;
    \item \textit{move}: all atoms move in parallel to their target positions, with duration determined by the maximum displacement;
    \item \textit{deactivate}: atoms transfer back to the static SLM traps.
\end{enumerate}
Increased depth necessitates more frequent atom transfers, which elevates the probability of atom loss and decoherence.
Moreover, physical \emph{execution duration} captures the total wall-clock time including both the rearrangements and gate operations, which is dominated by the physical displacement in each cycle. This duration directly governs the processor's clock rate and idling error accumulation during transport, making its minimization critical for high-fidelity, large-scale computation.

%% ---------------------------------------------------------------------------
\subsection{Structured Quantum Error Correction}
\label{sec:preli-qec}

Fault-tolerant quantum computing requires QEC to protect logical information from physical noise. A stabilizer code $\llbracket n, k, d \rrbracket$ encodes $k$ logical qubits into $n$ physical qubits with code distance~$d$ by specifying a set of \emph{stabilizer generators}
% ---multi-qubit Pauli operators whose eigenvalues reveal error syndromes without disturbing the encoded state 
\cite{gottesman1997,calderbank_quantum_1997}. Each round of \emph{syndrome extraction} measures every stabilizer via a sequence of two-qubit entangling gates between check and data qubits. 
% The cycle time of syndrome extraction directly governs the logical error rate, as longer cycles expose qubits to more decoherence.

QLDPC codes are a family of stabilizer codes in which each stabilizer acts on a bounded number of qubits and each qubit participates in a bounded number of stabilizers. This sparsity enables syndrome extraction with $\mathcal{O}(n)$ gates per round and makes these codes attractive for scalable QEC. Among them, HGP codes \cite{hgp,hgpcode} construct a quantum code from two classical seed codes $H_1 \!\in\! \mathbb{F}_2^{r_1 \times n_1}$ and $H_2 \!\in\! \mathbb{F}_2^{r_2 \times n_2}$ via the product:
\begin{equation}
\label{eq:hgp}
\mathcal{H}_X = \bigl[\, H_1 \!\otimes\! I_{n_2} \;\big|\; I_{r_1} \!\otimes\! H_2^{\mathsf{T}} \,\bigr], \quad
\mathcal{H}_Z = \bigl[\, I_{n_1} \!\otimes\! H_2 \;\big|\; H_1^{\mathsf{T}} \!\otimes\! I_{r_2} \,\bigr].
\end{equation}
The resulting code achieves high coding rate and large distance with bounded-weight stabilizers, offering favorable overhead scaling compared to surface codes \cite{xu_constant-overhead_2024}.

% The key structural property exploited by \Name is the decomposition of stabilizer interactions from the product structure. Because each block of $\mathcal{H}_X$ and $\mathcal{H}_Z$ in Eq.~\eqref{eq:hgp} acts along a single tensor factor, the two-qubit gates required for syndrome extraction naturally separate into two independent groups: one operating along the row dimension ($H_1$~factor) and one along the column dimension ($H_2$~factor). This row/column factorization is structurally compatible with the Cartesian AOD control geometry described in Section~\ref{sec:preli-na}, enabling the 2D compilation problem to decompose into independent 1D subproblems. This co-design opportunity is the foundation of the architecture protocol presented in Section~\ref{sec:overview}.

The decomposition can be read directly from Eq.~\eqref{eq:hgp}. Let the two HGP data-qubit blocks be indexed as $q^A_{j,\ell}$ for $(j,\ell)\!\in\![n_1]\!\times\![n_2]$ and $q^B_{i,m}$ for $(i,m)\!\in\![r_1]\!\times\![r_2]$; let $X$- and $Z$-check ancillas be indexed as $x_{i,\ell}$ and $z_{j,m}$. The nonzero entries induce exactly four edge types:
\begin{align*}
H_1(i,j)=1\Rightarrow\{(x_{i,\ell},q^A_{j,\ell}),(z_{j,m},q^B_{i,m})\}, \\
H_2(m,\ell)=1\Rightarrow\{(x_{i,\ell},q^B_{i,m}),(z_{j,m},q^A_{j,\ell})\}.
\end{align*}
The first two relations are copies of the $H_1$ Tanner graph at fixed $\ell$ or $m$, while the last two are copies of the $H_2$ Tanner graph at fixed $i$ or $j$. Thus every syndrome-extraction edge fixes one product coordinate and varies only the other; no edge has diagonal support. Placing these index sets as the four quadrants in Fig.~\ref{fig:intro}b maps the $H_1$ edges to horizontal 1D subproblems and the $H_2$ edges to vertical 1D subproblems. This exact edge partition, rather than a heuristic graph cut, is the structural guarantee that enables \Name to solve each 1D subproblem independently and compose the row/column phases on the Cartesian AOD geometry.

% The nonzero entries induce exactly four edge types: $H_1(i,j)=1$ gives $(x_{i,\ell},q^A_{j,\ell})$ and $(z_{j,m},q^B_{i,m})$, while $H_2(m,\ell)=1$ gives $(x_{i,\ell},q^B_{i,m})$ and $(z_{j,m},q^A_{j,\ell})$.

% The nonzero entries induce exactly four edge types:
% $H_1(i,j)=1\Rightarrow\{(x_{i,\ell},q^A_{j,\ell}),(z_{j,m},q^B_{i,m})\}$ and
% $H_2(m,\ell)=1\Rightarrow\{(x_{i,\ell},q^B_{i,m}),(z_{j,m},q^A_{j,\ell})\}$.

Related product-form code families, including general LP codes \cite{lpcode}, share analogous factorization structure. While we focus on HGP codes for concreteness and simplicity, \Name's approach has broader applicability to these codes to address similar underlying product structure as a critical subroutine, as discussed in Section~\ref{sec:lpcode} with a representative case study.

\section{Overview} \label{sec:overview}
This section provides an overview of \Name: the architecture protocol that enables layout synthesis on neutral-atom processors with a sequence of 1D execution plans, and the multi-phase optimization pipeline that produces optimized depth-optimal solutions.

\subsection{Architecture Protocol}

\begin{figure*}[t]
\centering
\includegraphics[width=0.66\linewidth]{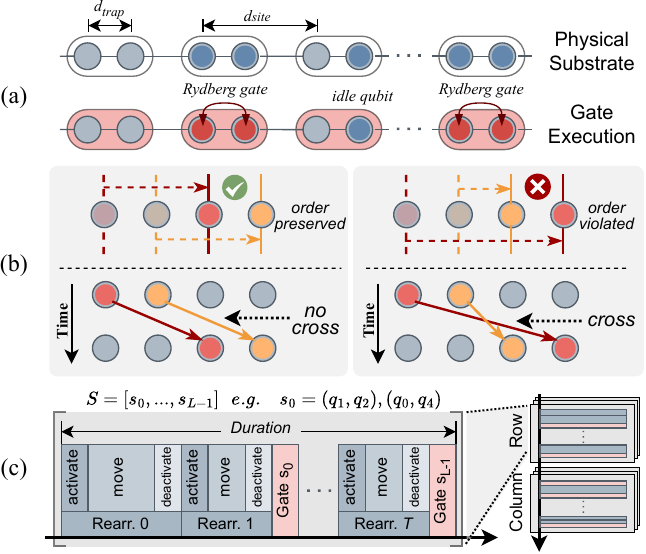}
% \caption{Architecture protocol of 1D physical execution. (a) 1D trap array and interaction model. (b) No-crossing constraints in 1D scenario. (c) Execution timeline.}
\caption{Architecture protocol of 1D physical execution.
(a)~1D interaction model: gate pairs co-locate for Rydberg execution, while idle qubits remain isolated.
(b)~No-crossing constraint: simultaneous AOD movements must preserve atom order.
(c)~Execution timeline: rearrangement steps are interleaved with scheduled gate stages and composed across rows/columns for product-code execution.}
\label{fig:arch-protocol}
\Description{}
\end{figure*}

Motivated by the native compatibility of product-structure qLDPC codes and orthogonal neutral-atom control, we decompose each product factor into a 1D atom execution plan with sequences of 1D rearrangements and gate operations along row or column dimensions. This dimensional simplification from 2D to 1D substantially alleviates the complexity of the constraint space. Combined with the $\mathcal{O}(n^2)$ reduction in problem size, this transformation enables the depth-optimal solution to each subproblem and defines our formal 1D execution protocol.

Fig.~\ref{fig:arch-protocol}a depicts the physical substrate and interaction model of our protocol. Qubits are trapped in a 1D array, where every pair of consecutive traps defines an interaction site. 
During the Rydberg blockade to achieve entanglement, any two qubits co-occupying the same site interact and perform a two-qubit gate.
The central no-crossing constraint of neutral-atom hardware, as detailed in Section~\ref{sec:preli-na}, is dimensionally simplified within our 1D protocol (Fig.~\ref{fig:arch-protocol}b). In contrast to the complex 2D coordination in prior studies \cite{atomique, enola, zac}, 1D rearrangement only requires that co-moving atoms do not bypass one another. Formally, any two atoms moving during the same time step must maintain their relative spatial order.

Fig.~\ref{fig:arch-protocol}c (left) illustrates the execution timeline. Given an input gate schedule (an ordered sequence of parallel gate stages), \Name produces a valid execution plan that specifies the atom rearrangements required for each time step interleaved with gate operations. After rearrangement steps bring the target qubits together, two-qubit gates are performed. The total execution duration is the sum of rearrangement and gate-operation durations, which is dominated by the former.
These derived 1D physical execution plans naturally extend to 2D architectures for product QEC codes through orthogonal decomposition. Specifically, the protocol executes all row-wise 1D rearrangements in parallel for one factor and then proceeds to the column-wise phase for the other, as aforementioned in Fig.~\ref{fig:intro}b and displayed in Fig.~\ref{fig:arch-protocol}c (right). 

Significantly, the flexibility of our protocol in qubit placement reveals a deeper architectural insight: rather than merely generating a rearrangement schedule, \Name actively explores the mapping from the QEC code to the physical hardware.
This formulation optimizes both the qubit-to-trap assignment and the resulting rearrangement, addressing the physical mapping and the logical interaction pattern as a single, unified planning problem. Such integrated co-optimization captures a vital advantage frequently neglected by prior works focused on isolated movement heuristics.

\begin{figure*}[t]
\centering
\includegraphics[width=0.95\linewidth]{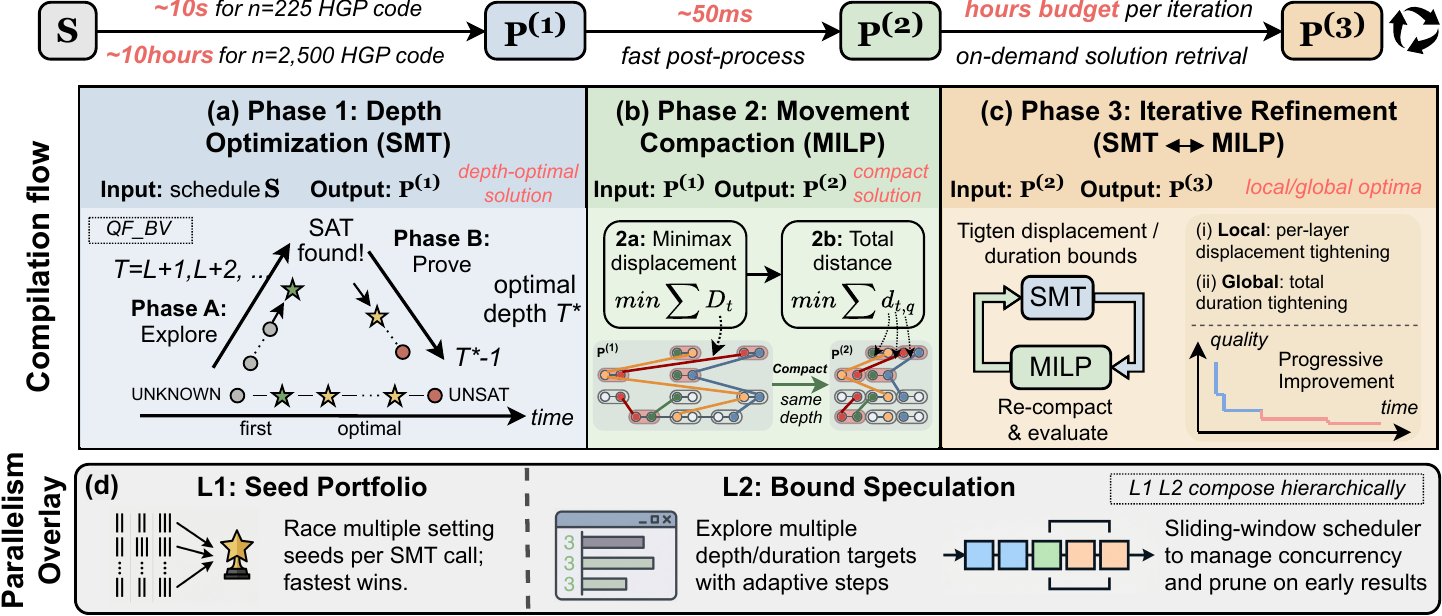}
\caption{Overview of the \Name compilation pipeline with runtime dataflow highlighted. 
(a) Depth optimization via SMT. (b) Movement compaction via MILP. (c) Iterative refinement with feedback loop. (d) Parallelism overlay.}
\label{fig:pipeline}
\Description{}
\end{figure*}

\subsection{Compilation Pipeline}

The duration of physical execution determines the clock rate of the neutral-atom processor. However, optimizing this duration involves a complex interplay of rearrangement depth, combinatorial topology, and physical displacement, making single-step optimization intractable. 
To address this challenge, \Name transforms a gate schedule as input into a fully optimized physical execution through three successive phases, each with a distinct optimization focus and leveraging different algorithmic techniques.
The input gate schedule is derived from edge coloring for the near-optimal circuit depth~\cite{enola}.
To further accelerate compilation, these phases are augmented with two-level parallelism to ensure practical utility.

Fig.~\ref{fig:pipeline} provides a complete view of this pipeline, with the three optimization phases in the main flow and the parallelism overlay below. 
Our multi-stage runtime profile highlights a progressive refinement strategy, where more expensive but fine-grained optimizations are executed in later phases, providing a robust foundation for anytime solving and on-demand solution retrieval.

\textbf{Phase 1: Depth optimization via SMT.} 
As the critical determinant of execution duration, we first optimize the depth $T$ of the protocol, defined as the total number of discrete time steps. To leverage the reasoning capabilities of modern SMT solvers, we formulate the physical execution planning as a quantifier-free bit-vector (QF\_BV) satisfiability instance \cite{z3, olsq_dpqa_2} and search for the minimal depth $T^*$ via bidirectional deepening. As shown in the inset of Fig.~\ref{fig:pipeline}a, this search proceeds in two directions: an upward exploration phase probes increasing depths for a rapid satisfying assignment, followed by a downward proof phase that verifies optimality by demonstrating the infeasibility of $T^*-1$. As a result, Phase 1 yields a provably depth-optimal execution plan that favorably maximizes the number of parallel rearrangement steps, reducing execution duration by more than 80\% relative to the baseline.
See Appendix~\ref{sec:base} for formulation details of Phase~1.
% Further details of Phase~1 are provided in Appendix~\ref{sec:base}.

\textbf{Phase 2: Movement compaction via MILP.} The depth-optimal solution maximizes parallelism but disregards physical distances. Phase 2 compacts trap-site assignments to minimize the maximum physical displacement per rearrangement step, and secondarily the total displacement across all qubits. Given that these objectives align perfectly with the strengths of mixed-integer linear programming (MILP) in minimax problems, we utilize an MILP formulation in Phase 2, as annotated in Fig.~\ref{fig:pipeline}b, delivering solutions with 11\% to 16\% tighter total displacement than the Phase~1 output per compaction. Crucially, the MILP preserves the discovered topology, thereby maintaining the depth optimality while significantly decreasing the complexity itself.
See Appendix~\ref{sec:compact} for details of Phase~2.

\textbf{Phase 3: Iterative refinement via feedback.} Phase 1 and 2 operate in a feedforward manner: the SMT solver finds one depth-optimal combinatorial topology, and the MILP compresses it. However, among the potentially vast space of depth-optimal topologies, some are inherently more amenable to compaction than others. Phase 3 closes this gap through the iterative feedback loop depicted in Fig.~\ref{fig:pipeline}c, which alternately (a) tightens physical displacement or duration bounds in the SMT formulation, and (b) re-compacts the solution via MILP. This refinement yields an additional duration reduction of up to 27\%.
Specifically, two strategies are employed: 
(1) \textit{Local duration tightening} identifies the current bottleneck rearrangement step and reduces its maximum displacement bound, terminating once no further improvement is possible;
(2) \textit{Global duration tightening} encodes the total duration directly into the SMT formulation, and adaptively tightens this bound until convergence. 
% Further details of Phase~3 are provided in Appendix~\ref{sec:feedback}.
See Appendix~\ref{sec:feedback} for more details of Phase~3.

% \begin{enumerate}[leftmargin=*]
%     \item \textit{Local duration tightening} identifies the current bottleneck rearrangement step and reduces its maximum displacement bound, terminating once no further improvement is possible. 
%     % This local search efficiently navigates the solution towards more compact topologies but can stall at a uniform distribution.
%     \item \textit{Global duration tightening} encodes the total duration directly into the SMT formulation, and adaptively tightens this bound until convergence. 
%     % This procedure extricates the solution from uniform-distribution local optima and provides a means to quantify the gap relative to global optimality.
% \end{enumerate}

% Main results table: HGP SE quality comparison (rebuttal)
% Same layout as v4, with Enola, ours, and delta columns highlighted.

\begin{table*}[t]
\centering
\setlength{\tabcolsep}{3.4pt}
\caption{High-rate HGP code demonstration. Comparison among Xu et al.~\cite{xu_constant-overhead_2024}, Enola~\cite{enola}, and ours. Clock Rate denotes the syndrome extraction frequency as $1/\text{duration}$ (Hz). For depth, $\Delta$ is baseline/ours; for clock rate, $\Delta$ is ours/baseline.}
\label{tab:hgp_results}
\resizebox{0.9\textwidth}{!}{%
\begin{tabular}{c|ccccc|ccc|ccccc}
\toprule
\multirow{2}{*}{\textbf{$n$}}
  & \multicolumn{5}{c|}{{\small \textbf{Exec.\ Depth}}}
  & \multicolumn{3}{c|}{{\small \textbf{Duration} (ms)}}
  & \multicolumn{5}{c}{{\small \textbf{Clock Rate} (Hz)}} \\
  & {\small Xu et al.} & {\small \textbf{$\Delta$}} & {\small \textbf{Enola}} & {\small \textbf{$\Delta$}} & {\small \textbf{Ours}}
  & {\small Xu et al.} & {\small \textbf{Enola}} & {\small \textbf{Ours}}
  & {\small Xu et al.} & {\small \textbf{$\Delta$}} & {\small \textbf{Enola}} & {\small \textbf{$\Delta$}} & {\small \textbf{Ours}} \\
\midrule
225
  & 96 & \boldmath{$8.0\times$} & 366 & \boldmath{$30.5\times$} & \textbf{12}
  & 13.34 & 88.13 & \textbf{2.20}
  & 74.97 & \boldmath{$6.1\times$} & 11.35 & \boldmath{$40.1\times$} & \textbf{455.23} \\
625
  & 108 & \boldmath{$6.8\times$} & 450 & \boldmath{$28.1\times$} & \textbf{16}
  & 17.13 & 133.90 & \textbf{4.50}
  & 58.39 & \boldmath{$3.8\times$} & 7.47 & \boldmath{$29.8\times$} & \textbf{222.46} \\
1225
  & 112 & \boldmath{$7.0\times$} & 568 & \boldmath{$35.5\times$} & \textbf{16}
  & 20.20 & 200.12 & \textbf{5.18}
  & 49.52 & \boldmath{$3.9\times$} & 5.00 & \boldmath{$38.6\times$} & \textbf{192.99} \\
2500
  & 126 & \boldmath{$7.0\times$} & 684 & \boldmath{$38.0\times$} & \textbf{18}
  & 24.39 & 280.36 & \textbf{6.66}
  & 41.01 & \boldmath{$3.7\times$} & 3.57 & \boldmath{$42.1\times$} & \textbf{150.10} \\
\bottomrule
\end{tabular}%
}
\end{table*}

\textbf{Parallelism overlay.} 
Coupled with the three-phase pipeline, \Name exploits two levels of parallelism to reduce compilation wall-clock time, summarized in Fig.~\ref{fig:pipeline}d. Seed portfolio parallelism (L1) races multiple random-seed variants of each SMT call to exploit the inherent runtime variance. Bound speculation parallelism (L2) explores multiple depth or duration targets simultaneously, managed via a sliding-window scheduler. This scheduler dynamically prunes candidates rendered irrelevant or infeasible by early results, thereby accelerating convergence. These schemes compose hierarchically, with each instance internally employing seed portfolio and bound speculation. Incremental solving is also leveraged to retain learned clauses across iterations for single-seed scenarios.
See Appendix~\ref{sec:parallel} for details of compilation-time parallelism.

% \section{Methodology} \label{sec:method}
% \input{sections/4-methodology}

\section{Evaluation} \label{sec:eval}

\subsection{Experimental Setup}
\Name is implemented in Python, with the core compilation pipeline built on top of two solver backends.
The SMT-based phases employ the Z3 solver (v4.16.0) \cite{z3}, 
invoked through its Python API. 
The MILP-based compaction phase is solved by HiGHS in SciPy (v1.14.1) \cite{scipy}.
% , a high-performance open-source mixed-integer linear programming solver. 
For QEC evaluation, we simulate syndrome extraction circuits using Stim (v1.15.0) \cite{gidney2021stim}, and decode syndromes with the belief-propagation plus ordered-statistics decoding (BP-OSD) decoder \cite{roffe_decoding_2020, Roffe_LDPC_Python_tools_2022}. All experiments are conducted on a server equipped with an AMD EPYC 9654 Processor at 2.4 GHz with 192 cores, with maximum 5GB of RAM allocated for each SMT instance.
% and 755 GB of RAM.

\textbf{Benchmarks.}
We evaluate \Name on two categories of benchmarks, with the same approach used in prior work~\cite{xu_constant-overhead_2024}.
For \emph{high-rate HGP code application}, we adopt existing hypergraph product codes with favorable parameters, spanning a range of code sizes from $\llbracket 225, 9, 4 \rrbracket$ to $\llbracket 2500, 100, 12 \rrbracket$.
For \emph{1D physical execution analysis}, we construct gate schedules with edge-coloring from randomly generated classical $(3,4)$-regular Tanner graphs, which define bipartite connectivity between variable and check nodes and are widely used for qLDPC code syndrome extraction patterns.

\textbf{Baseline.}
We compare against the state-of-the-art heuristic by Xu et al. \cite{xu_constant-overhead_2024}, which operates on arbitrary 1D rearrangements with $\mathcal{O}(\log n)$ scaling depth. 
Additionally, considering our architectural model here as a monolithic layout, we also compare against Enola~\cite{enola}, the leading compiler designed delicately for the 2D monolithic layout synthesis with similar scheduling. Baselines and our work are evaluated under the same neutral atom array architecture for a fair comparison, with $N$ interaction sites per row or column
for an $N$-qubit subproblem per dimension.
% where $N$ is the number of qubits for a fair comparison.
% where $N$ is the number of qubits to guarantee a solvable and fair comparison.

\textbf{Physical Model and Parameters.}
The rearrangement duration is obtained by the kinematic model introduced in Eq.~\eqref{eq:kinematic}. We adopt the physical parameters from~\cite{bluvstein_logical_2024, zac}: Rydberg entangling gate duration $t_{\text{gate}} {=} 0.36\,\si{\micro\second}$, trap-transfer time $t_{\text{transfer}} {=} 15\,\si{\micro\second}$, atom acceleration $\alpha {=} 2.75\times10^{-3}\,\si{\micro\meter/\micro\second^2}$, inter-site spacing $d_{\text{site}} {=} 12\,\si{\micro\meter}$, and intra-site trap spacing $d_{\text{trap}} {=} 2\,\si{\micro\meter}$.
The total syndrome extraction cycle duration is the primary compilation-quality metric and directly governs the clock rate and error budget for fault-tolerant operation. 
Specifically, following the approach in~\cite{xu_constant-overhead_2024}, we inherit the same circuit-level noise model and non-idling error channels, and model idling error accumulated during atom rearrangement as a linear approximation with gate error $p_g$:
% Specifically, we model idling error accumulated during atom rearrangement as a linear approximation with gate error $p_g$, following the approach in~\cite{xu_constant-overhead_2024}:
\begin{equation*}
    p_i = t_{\text{rearrange}}/T_c \times p_g / p_{0},
\end{equation*}
where $T_c=10\,\mathrm{s}$ is the atom coherence time and $p_{0}=0.005$ is the current CZ gate infidelity demonstrated in~\cite{evered_high-fidelity_2023}.

\subsection{Main Results}

\subsubsection{High-Rate HGP Code Application} \label{sec:hgp_app}

We apply \Name to compile the full syndrome extraction cycle of HGP codes mapped onto neutral atom arrays, ranging from $\llbracket 225, 9, 4 \rrbracket$ to $\llbracket 2500, 100, 12 \rrbracket$.
% Because the HGP product structure decomposes interactions into orthogonal row and column groups (Section~\ref{sec:preli-qec}), a complete syndrome extraction round requires one 1D rearrangement per row group plus one per column group. All metrics below account for this composition.

Table~\ref{tab:hgp_results} compares the key performance metrics for the high-rate HGP code demonstrations. 
Compared with Enola's general 2D compilation, \Name reduces execution depth by 28.1$\times$ to 38.0$\times$ across all code sizes.
Even under dimensional decomposition, \Name still achieves a 6.8$\times$ to 8.0$\times$ reduction over the asymptotic 1D constructive algorithm of Xu et al., bringing the syndrome extraction cycle down to single-digit milliseconds. The absolute duration gap also widens with code size, from 11.1\,ms at $n{=}225$ to 17.7\,ms at $n{=}2{,}500$, reflecting the growing advantage of \Name's near-constant-depth solutions over the baseline's logarithmically scaling depth.

Critically, the reduction in cycle duration directly translates to higher clock rates for the quantum processors, enabling a 3.7$\times$ to 6.1$\times$ improvement over the routing-centric algorithm of Xu et al. and 29.8$\times$ to 42.1$\times$ over Enola's general solutions.
These results confirm that co-optimizing atom rearrangement with global consideration and code property drives FTQC performance on neutral-atom architectures substantially, as rearrangements dominate over 99\% of the syndrome extraction duration here.
This product-aware co-optimization naturally generalizes beyond HGP codes, as illustrated by the representative case study later in Section~\ref{sec:lpcode}.

In terms of scaling, by decomposing row and column sub-problems, \Name ensures that computational costs depend only on the 1D problem size rather than the total qubit count. This decomposition enables the efficient compilation of systems exceeding 2{,}500 data qubits within hours (Section~\ref{sec:eval-scalability}), a runtime well within the acceptable range for one-time offline QEC processes.

\subsubsection{1D Physical Execution Analysis} \label{sec:1d_app}
In addition to the specific HGP code application, we also evaluate \Name on the 1D physical execution subroutine to analyze its performance across various scales and topologies.
Table~\ref{tab:main_results} summarizes the comparison among the Xu et al. baseline \cite{xu_constant-overhead_2024}, the leading 2D monolithic compiler Enola~\cite{enola}, and \Name across all five benchmark sizes.

% Main results table: 1D rearrangement quality comparison (v2)
% Adds per-baseline delta columns where delta denotes ours compared to each
% baseline. Since depth and duration are lower-is-better, delta is baseline/ours.

\begin{table*}[t]
\centering
\setlength{\tabcolsep}{5pt}
\caption{1D physical execution quality comparison among Xu et al.~\cite{xu_constant-overhead_2024}, Enola~\cite{enola}, and ours. $\Delta$ denotes baseline/ours.}
\label{tab:main_results}
\resizebox{0.75\textwidth}{!}{%
\begin{tabular}{c|ccccc|ccccc}
\toprule
\multirow{2}{*}{\textbf{Size}}
  & \multicolumn{5}{c|}{{\small \textbf{Exec.\ Depth} ($T$)}}
  & \multicolumn{5}{c}{{\small \textbf{Exec.\ Duration} (ms)}} \\
  & {\small Xu et al.} & {\small $\Delta$} & {\small Enola} & {\small $\Delta$} & {\small Ours}
  & {\small Xu et al.} & {\small $\Delta$} & {\small Enola} & {\small $\Delta$} & {\small Ours} \\
\midrule
14
  & 30.4 & \boldmath{$7.6\times$} & 14.4 & \boldmath{$3.6\times$} & \textbf{4.0}
  & 3.74 & \boldmath{$8.0\times$} & 2.05 & \boldmath{$4.4\times$} & \textbf{0.47} \\
21
  & 37.0 & \boldmath{$9.0\times$} & 18.2 & \boldmath{$4.4\times$} & \textbf{4.1}
  & 4.98 & \boldmath{$8.9\times$} & 3.06 & \boldmath{$5.5\times$} & \textbf{0.56} \\
28
  & 37.0 & \boldmath{$8.6\times$} & 21.9 & \boldmath{$5.1\times$} & \textbf{4.3}
  & 5.57 & \boldmath{$8.3\times$} & 4.34 & \boldmath{$6.5\times$} & \textbf{0.67} \\
35
  & 41.0 & \boldmath{$8.7\times$} & 24.5 & \boldmath{$5.2\times$} & \textbf{4.7}
  & 6.37 & \boldmath{$8.6\times$} & 5.39 & \boldmath{$7.3\times$} & \textbf{0.74} \\
42
  & 42.5 & \boldmath{$8.5\times$} & 29.2 & \boldmath{$5.8\times$} & \textbf{5.0}
  & 7.06 & \boldmath{$8.4\times$} & 7.31 & \boldmath{$8.7\times$} & \textbf{0.84} \\
\midrule
\multicolumn{1}{c|}{\textbf{$\Delta$ Avg.}}
  & --- & \boldmath{$8.5\times$} & --- & \boldmath{$4.8\times$} & ---
  & --- & \boldmath{$8.4\times$} & --- & \boldmath{$6.5\times$} & --- \\
\bottomrule
\end{tabular}%
}
\end{table*}

\begin{figure*}[t]
\centering
\includegraphics[width=0.8\linewidth]{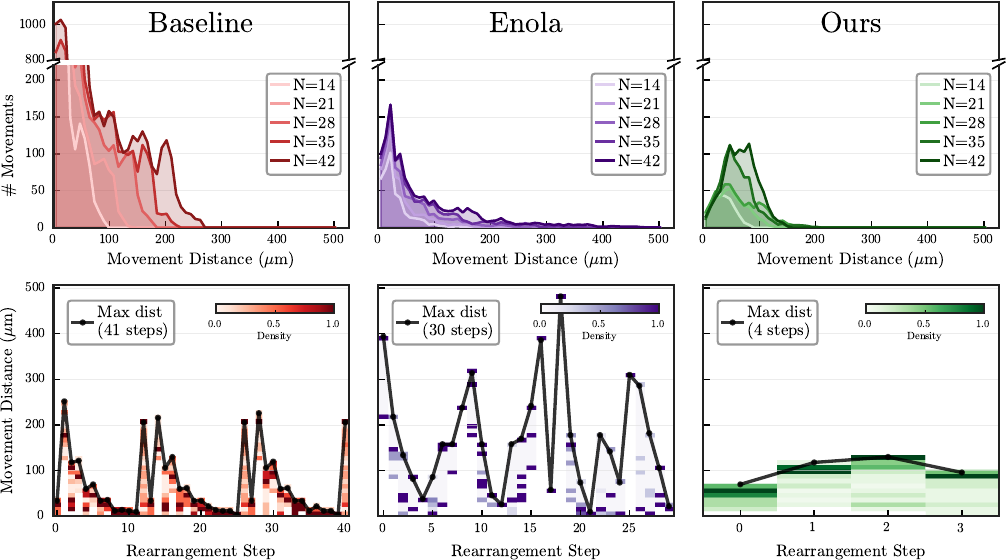}
\caption{Movement profile for the Xu et al. baseline, Enola, and ours on 1D execution benchmarks. Top row: distribution of individual atom displacement distances across all problem sizes. Bottom row: per-step movement profile for a representative $N{=}42$ instance, where color intensity indicates the density of movements at a given distance and the black envelope traces the maximum displacement per step.}
\label{fig:movement_analysis}
\Description{}
\end{figure*}

\textbf{Execution depth.}
Minimizing the number of rearrangement steps $T$ is the essential objective in our compilation. Fewer steps imply higher parallelism, enabling more simultaneous atom movements and thereby reducing the overall execution duration. Conversely, more steps increase atom transfers, resulting in higher atom loss error accumulation and greater transfer time overhead. 

The Xu et al. constructive algorithm produces 30 to 43 steps, growing as $\mathcal{O}(\log N)$ with problem size. Enola achieves 14 to 29 steps via separated placement and routing compilation.
\Name achieves a near-constant depth of 4--5 steps across all tested sizes, representing an 86.8--88.9\% reduction over Xu et al.
This near-constant depth is produced by \Name's SMT formulation with global consideration, which discovers maximally parallel topologies that fully unleash the potential for simultaneous movements.
Notably, the 1D subroutine shows an even wider performance margin than the full HGP application, where multi-round syndrome extraction imposes strict cyclic scheduling constraints that we address with a lightweight solver for returning to the initial placement.

% To complete the cycle, our mapping currently simply reverses the 1D subroutine to its initial state, with further performance gains expected from a more sophisticated strategy.

% The result indicates that, for the structured gate schedules arising in QEC syndrome extraction, the combinatorial lower bound $T_{\text{lb}} = L$ (number of gate stages) is nearly tight at relatively small problem size.

\begin{figure*}[t]
\centering
\includegraphics[width=0.8\linewidth]{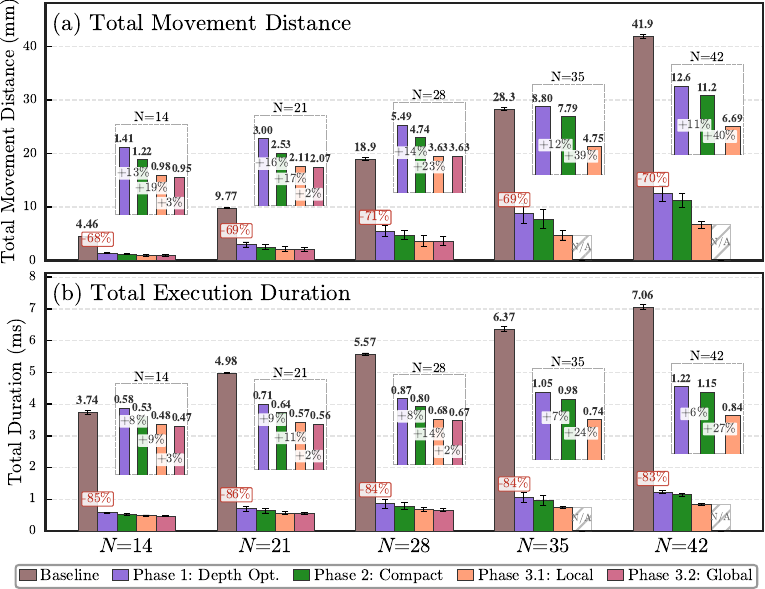}
% \caption{Pipeline stage contribution: (a) total movement distance, and (b) execution duration. Main bars compare the baseline with each pipeline stage across five problem sizes. Inset panels zoom into the pipeline stages (S1--S3.2) with per-stage improvement annotations.}
\caption{Pipeline stage contribution: (a) total movement distance, and (b) execution duration. Main bars compare the baseline with each pipeline stage across five problem sizes. Inset panels zoom into per-stage improvement.}
\label{fig:exp_pipeline}
\Description{}
\end{figure*}

\textbf{Execution duration.}
As the primary performance metric, execution duration determines the interval between operations and directly dictates the clock rate of the neutral-atom system.
Across all sizes, \Name achieves an 8.0$\times$ to 8.9$\times$ speedup over the Xu et al. baseline and a 4.4$\times$ to 8.7$\times$ speedup over Enola. This reduction translates directly into higher clock rates, addressing the architectural bottleneck and improving execution efficiency. Moreover, the reduction in duration mitigates both idling noise and atom loss, thereby enhancing overall system fidelity.
% Moreover, although the depth reduction alone accounts for the majority of the duration improvement, the distance compaction in Phases~2--3 contributes additional reduction, as we detail in the ablation study (Section~\ref{sec:eval-ablation}).

This significant duration reduction stems from both depth reduction and compaction of movement distances. Fig.~\ref{fig:movement_analysis} illustrates this via displacement distributions (top row) and per-step movement profiles (bottom row) for a representative instance. 

The top row reveals a clear contrast in how each approach distributes displacement. The Xu et al. baseline is characterized by a high volume of short-distance movements below $100\,\mu$m, and Enola reduces the number of movements but suffers from a heavy tail beyond $400\,\mu$m. In contrast, \Name maintains a compact distribution that remains tightly concentrated below $150\,\mu$m with no heavy tail. 

The bottom row exposes why depth reduction alone can be insufficient. Although Enola reduces depth from 41 to 30 steps, its maximum per-step displacement jumps from $252\,\mu$m to $482\,\mu$m and its critical-path displacement indicates these long-range shuttles are not well organized to overlap, offsetting the depth savings entirely. By comparison, compressing the schedule to 4 steps with a controlled maximum displacement, \Name achieves a total duration 7.8$\times$ better than Xu et al. and 8.8$\times$ better than Enola, demonstrating the effectiveness of our joint depth-displacement optimization.

\subsection{Ablation Study}
\label{sec:eval-ablation}
\subsubsection{Pipeline Stage Contribution}
To isolate the contribution of each compilation phase, we evaluate the intermediate outputs at every stage of the \Name pipeline: Phase~1 (SMT depth optimization), Phase~2 (MILP movement compaction), Phase~3.1 (local duration tightening), and Phase~3.2 (global duration tightening).

\textbf{Displacement compaction.}
Fig.~\ref{fig:exp_pipeline}a presents the total movement distance after each pipeline stage. Phase~1 already achieves a substantial reduction over the Xu et al. baseline (68--71\%) by virtue of the depth-optimal topology, which eliminates unnecessary long-range shuttles inherent in the baseline's recursive strategy. Phase~2 contributes an additional 11--16\% reduction by re-assigning positions within the fixed topology to minimize per-cycle displacement. Phase~3.1 provides a further 17--40\% improvement, particularly at larger sizes where more transitions have room for tightening. 
% Phase~3.2 (global search) adds a modest 0--3\% to the achievable optimum within the time budget, with the challenge of pre-computation and encoding duration becoming more significant at larger scales.

The inset panels of Fig.~\ref{fig:exp_pipeline}a zoom into the pipeline stages for each problem size, annotating the incremental percentage reduction at each step. A clear trend emerges: the MILP compaction and local feedback are complementary; the former optimizes within a fixed topology, while the latter steers the SMT solver toward topologies that are inherently more compact.

\begin{figure*}[t]
\centering
\includegraphics[width=0.85\linewidth]{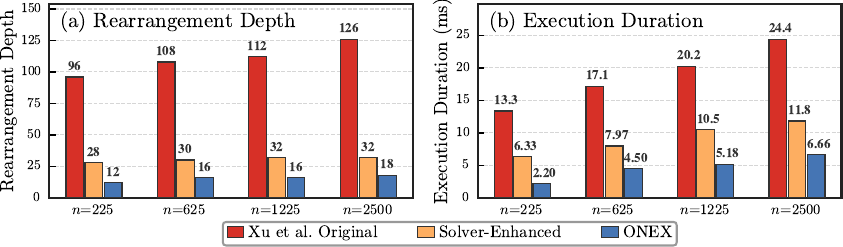}
\caption{
Formulation contribution breakdown: (a) rearrangement depth, and (b) execution duration. Bars compare Xu et al.'s constructive baseline, a solver-enhanced baseline w/o global formulation, and ONEX across four HGP codes.
}
\label{fig:exp_formulation}
\Description{}
\end{figure*}

\begin{figure*}[t]
\centering
\includegraphics[width=0.88\linewidth]{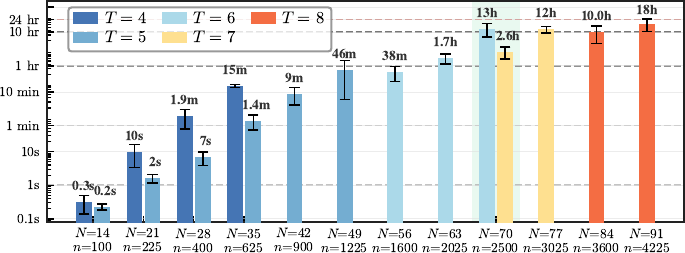}
\caption{Phase 1 solving time versus problem size $N$ (lower label: corresponding HGP data qubit counts~$n$); bars within a group represent different execution depths~$T$. 
Dashed lines mark reference time budgets.
}
\label{fig:exp_scalability}
\Description{}
\end{figure*}

\textbf{Duration progression.}
Fig.~\ref{fig:exp_pipeline}b shows the total physical execution duration following the same decomposition. 
The qualitative pattern mirrors the observed reduction in movement distance, though the square-root model modulates the specific impact on duration. Given that the duration is sensitive to the longest movement in each step, the compatible improvement to total distance reduction indicates that \Name successfully optimizes the bottleneck movement rather than merely improving average case metrics.

Experimental results demonstrate that the global search in Phase 3.2 contributes only a marginal reduction in both distance and duration. This outcome validates the effectiveness of the preceding local search, which maintains a narrow optimality gap relative to the global optimum. Furthermore, the limited gains with high runtime overhead highlight the inherent challenges of direct global optimization at larger scales, thereby confirming that our efficient feedforward approach is practical and crucial.

\subsubsection{Formulation Breakdown}
Beyond the compilation pipeline itself, we further isolate the contribution of \Name's formulation under the same dimensional decomposition. 

To this end, we construct a solver-enhanced baseline on top of Xu et al. by replacing its constructive algorithm with the SMT and MILP solving, while retaining its local, routing-centric formulation. In contrast, \Name treats compilation as a global optimization problem: it jointly explores qubit mapping, placement, and rearrangement to optimize the complete execution plan.

Fig.~\ref{fig:exp_formulation} presents this formulation-level breakdown in solution quality. The solver-enhanced baseline improves over the original baseline by 3.43$\times$--3.94$\times$ in depth and 1.93$\times$--2.15$\times$ in duration, demonstrating the benefit of locally optimal solving alone. \Name further improves over this enhanced baseline by 1.78$\times$--2.33$\times$ in depth and 1.77$\times$--2.88$\times$ in duration. This multiplicative breakdown shows that both the solver-based optimization and the global formulation make substantial, complementary contributions to the final solution quality.

\subsection{Scalability Analysis}
\label{sec:eval-scalability}
We analyze the scalability based on two experiments: compilation time as a function of problem size and the convergence behavior of solution quality during the feedback phases. 
Since MILP-based compaction consistently finishes in under 100\,ms, our scalability analysis focuses on Phase 1 solving time and the feedback process.

\textbf{Compilation time versus problem size.}
Fig.~\ref{fig:exp_scalability} reports the Phase 1 solving time to obtain the feasible solution at different depths across 1D problem sizes from $N{=}14$ to $N{=}91$, corresponding to HGP codes with 100 to 4{,}225 qubits.
As expected for an NP-hard combinatorial formulation, compilation time grows exponentially with problem size.
% At small scales ($N \leq 28$), \Name finds depth-optimal solutions ($T{=}4$--$5$) in under 2~minutes, well within an interactive compilation workflow.
The practical compilation boundary lies near $N{=}70$ (HGP~$n{=}2{,}500$), where depth-$T{=}6$ solutions require a mean of 12.8~hours; at $N{=}91$ (HGP~$n{=}4{,}225$), non-speculative compilation averages 17.6~hours for $T{=}8$. However, given that QEC compilation is a one-time offline task, these time scales are acceptable to achieve superior solution quality. In practical applications, the anytime solving property and speculation parallelism of \Name also temper this exponential growth, facilitating a fast first-solution turnaround.

\begin{figure*}[t]
\centering
\includegraphics[width=0.88\linewidth]{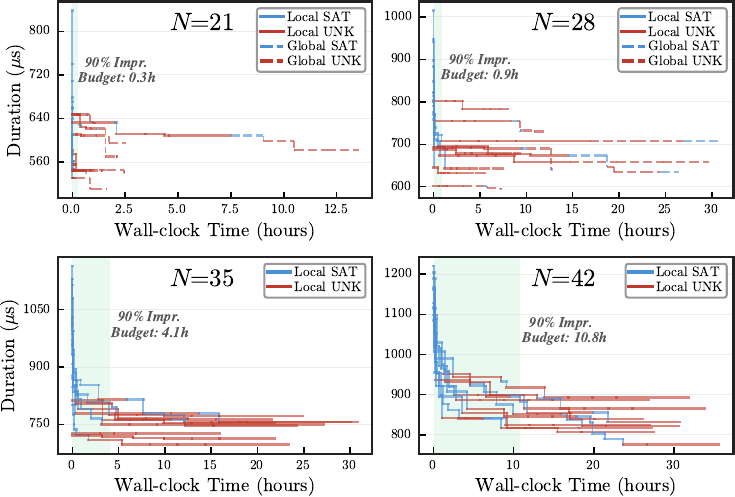}
\caption{Solution quality evolution during the feedback phases with solver outcome decomposition across four problem sizes ($N{=}21$ to~$42$). Segment color encodes the solver outcome (blue: sat, red: unsat/timeout); line style encodes the search phase (solid: local, dashed: global). The green shaded region marks the per-size 90\% improvement budget.}
\label{fig:exp_feedback}
\Description{}
\end{figure*}

\textbf{Solution quality versus compilation time.}
Fig.~\ref{fig:exp_feedback} shows the physical execution duration as a function of wall-clock time during the iterative feedback phases across four benchmark sizes, decomposing each iteration segment by its solver outcome.

The convergence profiles exhibit a characteristic diminishing-returns shape. Local search captures the majority of improvement early with a sharp initial drop in the feedback process. Subsequent iterations encounter increasingly tight bounds that resolve as unsat or timeout, yielding progressively smaller gains. Moreover, global search only provides minor additional benefit primarily at smaller sizes, where the solver completes enough iterations within its time budget to explore the bound space effectively. 

This convergence behavior further justifies the time management of the anytime strategy, which helps to capture the most significant improvements early while avoiding the long tail of the optimization plateau. 
Specifically, we calculate the 90\% improvement budgets (green shaded region), offering a practical balance between solution quality and compilation time. These budgets achieve 90\% of the total gain with only a moderate fraction of the time, spanning from 0.3 hours at $N{=}21$ to 10.8 hours at $N{=}42$.

\section{Architecture Adaptation: Zoned Layouts} \label{sec:zoned}
Beyond the monolithic architecture used in the main evaluation, ONEX also provides a natural path toward zoned architectures~\cite{bluvstein_logical_2024},
aligning with the current trajectory of neutral-atom hardware.

\begin{figure*}[t]
\centering
\includegraphics[width=0.82\linewidth]{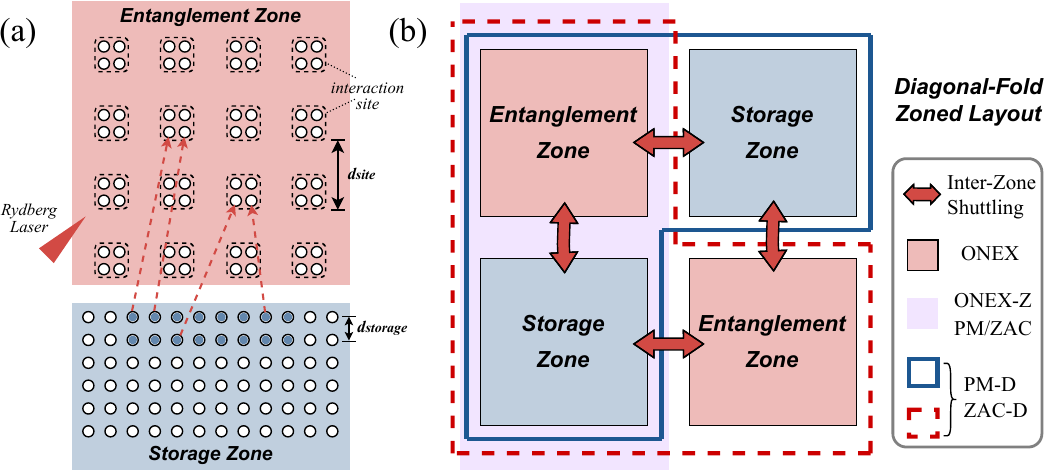}
\caption{Zoned architecture for neutral-atom quantum computing. (a) Distinct regions for entanglement and storage. (b) Layouts for different compilation strategies.}
\label{fig:zone}
\Description{}
\end{figure*}

Fig.~\ref{fig:zone}a illustrates the typical zoned architecture, featuring distinct regions for entangling gates and qubit storage. The Rydberg laser is restricted to the entanglement zone to activate qubits in the same interaction site for gate operations, while qubits in the storage zone remain densely packed to maximize capacity. 

\begin{figure*}[t]
\centering
\includegraphics[width=0.8\linewidth]{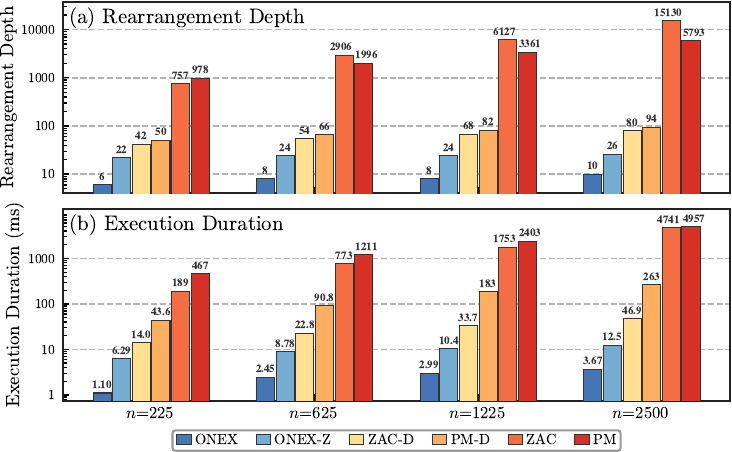}
\caption{Zoned-architecture adaptation: (a) rearrangement depth, and (b) execution duration. Both y-axes use logarithmic scale; dashed lines mark decade reference levels.}
\label{fig:exp_zone}
\Description{}
\end{figure*}

Fig.~\ref{fig:zone}b depicts the hardware layouts for different compilation strategies in our exploration. With the aligned physical substrate, \Name's execution plans can be directly applied to the entanglement zone without the need for the storage zone, so we preserve the original \Name notation for this strategy. To adapt \Name to the full zoned architecture and leverage the benefits of the storage zone, we also implement a zoned execution strategy, \Name-Z, which maps the rearrangements to the dense storage zone and incorporates additional loading and unloading steps to shuttle qubits between the storage and entanglement zones for gate execution. 

As baselines, we consider existing state-of-the-art compilers for zoned architectures, PowerMove~\cite{powermove} (PM) and ZAC~\cite{zac}, which rely on heuristic methods to generate valid circuit execution plans by rearranging qubits between zones for gate execution. In addition to their original general 2D compilation flows, we apply the same decomposition strategy used by \Name to these baselines, denoted as PM-D and ZAC-D, to isolate the impact of dimensional simplification. Under this decomposition, we propose a new diagonal-fold zoned layout for natively mapping their solutions, where each type of zone is placed along the diagonal to provide symmetric access to both directions and directly support the resulting 1D solutions.

Fig.~\ref{fig:exp_zone} compares the aforementioned zoned strategies across problem sizes. Since PM and ZAC do not natively support multi-round rearrangement, results are collected on single-round syndrome extraction circuits.
Compared with the general 2D zoned compilation flows of PM and ZAC, \Name achieves substantial speedups, reducing duration by over 1,000$\times$ when the problem size reaches $n{=}2{,}500$. By leveraging the decomposition strategy together with the diagonal-fold layout, PM-D and ZAC-D also achieve significant improvements over their original versions, confirming the effectiveness of dimensional simplification and delicate layout design in zoned scenarios. Notably, \Name still outperforms PM-D and ZAC-D by 37.1$\times$--71.5$\times$ and 9.3$\times$--12.8$\times$, respectively, demonstrating that \Name's entanglement-zone-only execution can still produce superior solutions.

\begin{figure*}[t]
\centering
\includegraphics[width=0.76\linewidth]{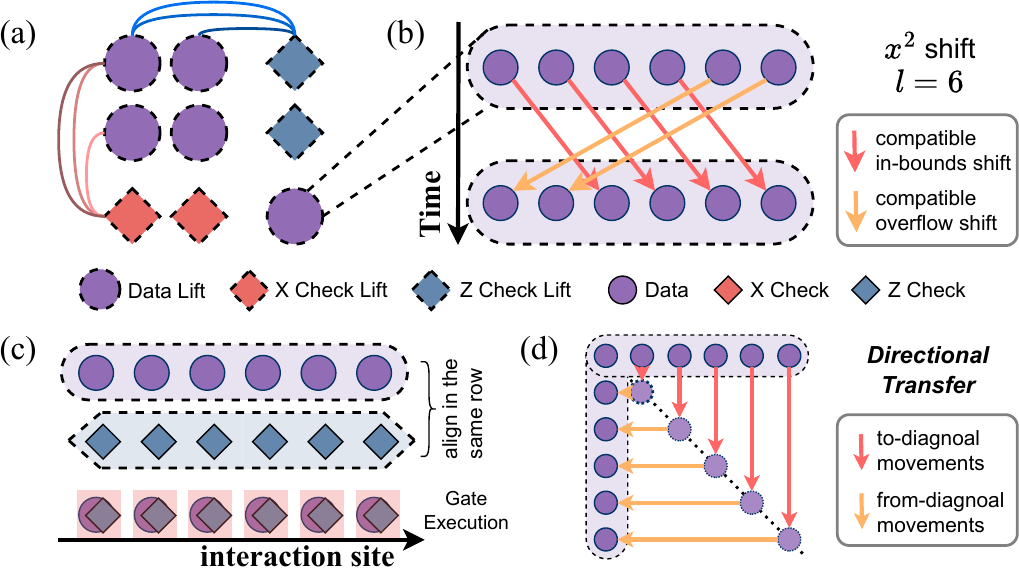}
\caption{LP-code lifting and product-aligned execution. (a)~An inter-lift interaction graph connects lifted data, $X$-check, and $Z$-check qubits. (b)~Each polynomial entry $x^k$ induces a cyclic shift among $\ell$ lifted copies. (c)~Lifted data and check qubits are aligned at interaction sites for parallel gate execution. (d)~Diagonal parking enables directional transfer between row- and column-oriented layouts.}
\label{fig:lpcode}
\Description{}
\end{figure*}

The comparison between \Name and \Name-Z provides direct architectural insight into the trade-off between intra-zone rearrangement and inter-zone shuttling. \Name-Z leverages high-density storage zones for intra-zone rearrangement, reducing movement distance by more than 50\% compared with \Name. However, the additional loading and unloading steps required to shuttle qubits between zones introduce significant overhead, increasing the total duration by 3.4$\times$--5.7$\times$ over \Name.
This overhead stems from two intrinsic factors. First, syndrome extraction circuits here have a high density of gate operations, which require frequent shuttling between zones, especially under our depth-optimal solutions. Second, the decomposition induces a tile-based placement to maximize parallelism, limiting the opportunity to group qubits near zone interfaces and further amplifying the shuttling cost. We expect workloads with lower gate density to benefit more from compact intra-zone rearrangement, even after accounting for inter-zone shuttling overhead. The fidelity impact of zoned execution is beyond the scope of this work and is left for future study.

\section{Code Generalization: Lifted Product} \label{sec:lpcode}
To further demonstrate the generality and extensibility of \Name beyond HGP codes, we deliver an example solution for a specific LP code. The example uses the $\llbracket 2610, 744, d\leq 16 \rrbracket$ LP code \eqref{eq:lp_code} from \cite{cain2026possible10000} with lift size $l=45$, which has been proposed as a promising high-rate memory code to enable practical applications for Shor's algorithm. To the best of our knowledge, no concrete execution strategy for this code has been presented before.

\begin{equation}
\label{eq:lp_code}
A =
\begin{pmatrix}
x^{29} & x^{21} & x^{31} & x^{15} & x^{37} & x^{25} & x^{27} \\
x^{13} & x^{25} & x^{19} & x^{26} & x^{11} & x^{18} & x^{29} \\
x^{31} & x^{2}  & x^{27} & x^{32} & x^{41} & x^{41} & x^{18}
\end{pmatrix}
\end{equation}

Briefly, our execution framework for general LP codes comprises four phases, illustrated through the above example in Fig.~\ref{fig:lpcode}:
\begin{enumerate}[label=(\alph*), widest=d, leftmargin=*]
% \begin{enumerate}[leftmargin=*]
    \item \textbf{Inter-lift rearrangement.} Starting from the seed base matrix, we treat each data or check lift as a scheduling unit and obtain a schedule through edge coloring. \Name is then applied to these units to determine the meta-level movements, which constitute the inter-lift rearrangement of the LP code.

    \item \textbf{Intra-lift rearrangement.} Within each active data lift, we use the corresponding polynomial elements of the base matrix to perform heuristic two-dimensional cyclic movements, thereby generating an intra-lift rearrangement. For the example LP code \eqref{eq:lp_code}, the polynomial elements are just monomials and can therefore be mapped directly to one-dimensional shifts.

    \item \textbf{Gate execution.} Following the inter- and intra-lift rearrangements, all data and check qubits are aligned at their designated positions for gate execution. Gates within compatible groups can then be executed in parallel, as in HGP demonstrations.

    \item \textbf{Directional transfer and phase repetition.} After completing execution along one dimension, we transfer the layout to the other dimension and repeat the preceding three phases. In the example considered here, we use a two-layer transfer scheme with diagonal parking to adapt the full layout between the horizontal and vertical dimensions.
\end{enumerate}
This framework reflects the theoretical construction of lifted product by addressing the inter- and intra-lift structure hierarchically. Together with the decomposition across product dimensions, it enables a feasible execution plan for general LP codes.

Building on the above framework, we obtain a concrete execution pipeline for the $\llbracket 2610, 744, d\leq 16 \rrbracket$ LP code, with performance estimates summarized in Table~\ref{tab:lpcode}.
Unlike dense HGP syndrome extraction, the repeated intra-lift rearrangements in this LP-code example lower the effective gate-execution density. Consequently, zoned execution reduces the intra-lift and inter-lift rearrangement times by approximately 46\% and 28\%, respectively. Even after accounting for the additional inter-zone shuttling overhead, ONEX-Z reduces the total cycle duration by approximately 10\%.

\begin{table}[t]
\centering
\setlength{\tabcolsep}{3pt}
\caption{Estimated execution-time breakdown for the $\llbracket 2610, 744, d\leq 16 \rrbracket$ LP code. All values are in milliseconds.}
\label{tab:lpcode}
\resizebox{\linewidth}{!}{%
\begin{tabular}{lcccccc}
\toprule
\textbf{Method} & \textbf{Total} & \textbf{Inter-lift} & \textbf{Intra-lift} & \textbf{Gate} & \textbf{Transfer} & \textbf{Shuttle} \\
\midrule
ONEX   & 23.731 & 10.573 & 11.280 & 0.005 & 1.873 & \textbf{0} \\
ONEX-Z & \textbf{21.405} & \textbf{7.560} & \textbf{6.060} & 0.005 & \textbf{0.996} & 6.783 \\
\bottomrule
\end{tabular}%
}
\end{table}

Notably, \Name{} remains a critical execution backbone of this framework by providing the core lift-level execution plan. Owing to the hierarchical construction of LP codes, each isolated lift-level subproblem is substantially smaller than that of an HGP code with the same size. This reduced problem size further alleviates scaling pressure on \Name while enabling high-quality execution plans. Moreover, this exploration demonstrates that the constituent units of \Name can themselves represent intricate two-dimensional patterns rather than just single qubits. This perspective also creates opportunities to integrate more sophisticated intra-lift designs, which we discuss further in Section~\ref{sec:discussion}.

\section{Discussion} \label{sec:discussion}
\textbf{Integration with Diverse Code Structures.}
\Name is designed to exploit the outer product structure shared by a wide range of qLDPC codes. Its product-structure-oriented optimization is therefore orthogonal to techniques that leverage inner code structure, such as cyclic symmetry or group-algebraic regularity~\cite{qsieve, Viszlai_2025,wang_coprime_2026}. These methods operate at different levels of the code hierarchy and do not conflict in their optimization targets. A natural direction for future work is therefore to compose product-structure decomposition with inner-structure-aware compilation. Such a compositional approach would extend the applicability to broader code families while preserving the benefits of each constituent technique.

A favorable inner-code structure is characterized by both a dense layout and efficient cyclic rearrangements. Taking the LP code in Section~\ref{sec:lpcode} as an example, qubits within each lift interact along only one direction and therefore map naturally to a linear layout rather than a compact two-dimensional patch. This mapping increases inner-lift displacement and incurs substantial duration overhead. In contrast, codes admitting dense patches can use the available space more efficiently. Additionally, shift-based cyclic rearrangements for a linear layout can be implemented using two opposing parallel movements, whereas more complex two-dimensional cycles may require additional heuristics or AOD resources to preserve efficient rearrangements among compatible groups.

% \Name is designed to exploit the product structure of a wide range of qLDPC codes; therefore the magnitude of the resulting benefit depends directly on the depth and dominance of this product structure in a given code family. For codes such as HGP codes, where the product construction fully determines the stabilizer interaction pattern, \Name captures the maximal parallelism and achieves the largest improvements. For code families where the product structure is shallow relative to additional algebraic structure (e.g., BB codes \cite{bbcode}, whose performance is dominated by the internal cyclic lift rather than the outer product), the decomposition provides a less significant fraction of the total optimization opportunity.

% Importantly, \Name's product-structure-oriented optimization is orthogonal to techniques that exploit inner code structure, such as cyclic symmetry or group-algebraic regularity \cite{qsieve, Viszlai_2025,wang_coprime_2026}. These methods operate at different levels of the code hierarchy and do not conflict in their optimization targets. A natural direction for future work is therefore to compose product-structure decomposition with inner-structure-aware compilation. Such a compositional approach would extend the applicability to broader code families while preserving the benefits of each constituent technique.

\textbf{Generalization to Fault-Tolerant Logic.}
Fault-tolerant quantum computation requires not only memory but also logic operations, such as lattice surgery and transversal gates, to manipulate logical information. Many of these operations, particularly surgery protocols for some product code families, retain the underlying product structure in their gate interaction patterns \cite{xu_fast_2025,chang_constant-time_2026,webster_explicit_2025}.
Moreover, logic layouts are typically less regular than memory layouts, lacking the common cyclic structure that admits trivial parallelism in syndrome extraction. This irregularity increases the gap between naive and optimized execution, presenting precisely the scenario where \Name's solver-based approach provides the greatest leverage. Extending \Name to compile logic operations would therefore enable a more complete characterization of the fault-tolerant compilation stack and demonstrate the framework's utility across the full spectrum of FTQC architectural primitives.

% The application evaluation in this work focuses on syndrome extraction circuits for QEC memory, where the mapping from stabilizer structure to physical execution is direct and well-defined. However, fault-tolerant quantum computation requires not only memory but also logic operations, such as lattice surgery and transversal gates, to manipulate logical information. Many of these operations, particularly surgery protocols for some product code families, retain the underlying product structure in their gate interaction patterns \cite{xu_fast_2025,chang_constant-time_2026,webster_explicit_2025}.

% Moreover, the physical execution challenge in logic operations may be more pronounced than in memory. Surgery layouts are typically less regular than memory layouts, lacking the simple cyclic structure that admits trivial parallelism in syndrome extraction. This irregularity increases the gap between naive and optimized execution, presenting precisely the scenario where \Name's solver-based approach provides the greatest leverage. Extending \Name to compile logic operations would therefore enable a more complete characterization of the fault-tolerant compilation stack and demonstrate the framework's utility across the full spectrum of FTQC architectural primitives.

\textbf{Versatility for Near-Term Architectures.}
While developed for fault-tolerant QEC, our protocol generates physical execution plans directly from gate scheduling, making it broadly applicable to general quantum circuit compilation. Its efficiency in managing dozens of qubits makes it particularly well-suited for the smaller-scale circuits of noisy intermediate-scale quantum (NISQ) devices. Moreover, since many quantum algorithms require high-count measurement shots, our 1D protocol enables parallel execution by duplicating circuits across multiple rows. 
This native support for compatible physical execution significantly reduces total hardware runtime, a critical advantage for practical utility of NISQ applications.

\section{Conclusion} \label{sec:conclusion}
% Scaling fault-tolerant quantum computing with high-rate qLDPC codes on neutral atom arrays requires efficient compilation of physical execution plans, a task whose combinatorial complexity grows rapidly with code size. In this work, we proposed \Name, a co-design framework that bridges the algebraic structure of product qLDPC codes with the Cartesian control geometry of atom rearrangement. 
% By decomposing the 2D physical execution planning into orthogonal 1D subproblems,
% % aligned with the natural factorization of stabilizer interactions, 
% \Name enables depth-optimal solutions for each subproblem with explicit SMT encodings.  
% Our three-phase compilation pipeline, combining depth optimization, MILP-driven movement compaction, and iterative feedback refinement, progressively improves solution quality with on-demand solution retrieval, while multi-level parallelism strategies maintain practical compilation times.
% Experimental evaluation on HGP codes with up to 2{,}500 data qubits demonstrates that \Name achieves syndrome extraction with a 70\% to 84\% reduction in cycle duration, indicating compatible 3.3$\times$ to 6.1$\times$ clock rates and 1.2$\times$ to 1.6$\times$ improvement in logical error rate. Looking ahead, extending \Name to logic operations and composing it with inner-code-aware techniques would broaden its applicability across diverse code families and the full fault-tolerant compilation stack.

Scaling fault-tolerant quantum computing with high-rate qLDPC codes on neutral atom arrays requires efficient compilation of physical execution plans. In this work, we proposed \Name, a structure-aware compilation framework for qLDPC codes with dimension-reduction properties. 
By exploiting the product structure and decomposing the 2D physical planning into orthogonal 1D subproblems, \Name enables optimized depth-optimal solutions through a practical multi-stage compilation pipeline.
Across HGP memory benchmarks with up to 2{,}500 data qubits, \Name substantially reduces syndrome extraction cycle duration, yielding clock rates 3.7$\times$--6.1$\times$ and 29.8$\times$--42.1$\times$ higher than those of the existing 1D constructive algorithm and general 2D compiler, respectively. Moreover, its adaptation to zoned layouts reveals the associated architectural trade-offs. Finally, a representative case study further extends the framework to the broader LP code family.

\begin{acks}
This work is supported by the U.S. National Science Foundation under Award No.~2533041 for NQVL:QSTD Phase~II—ORAQL: Open-Stack Rydberg Atom Quantum Computing Laboratory, a Phase~II continuation of DLPQC (Award No.~2410716), and under Award No.~2016245 for the Challenge Institute for Quantum Computation.
\end{acks}

\clearpage
\appendix
\section{Algorithm Details}
This section presents the formal problem formulation and the three successive optimization phases of \Name's compilation pipeline. We describe the SMT encoding for depth-optimal search (Section~\ref{sec:base}), the MILP formulation for movement compaction (Section~\ref{sec:compact}), the iterative feedback strategies for duration refinement (Section~\ref{sec:feedback}), and the multi-level parallelism overlay (Section~\ref{sec:parallel}). 

\subsection{Depth-Optimal SMT Encoding} \label{sec:base}

\subsubsection{Problem Statement}

Given $N$ qubits $\mathcal{Q}$ on $M$ traps across $T$ discrete time steps, and a gate schedule $\mathbf{S} = [s_0, \ldots, s_{L-1}]$ of $L$ stages, where each stage $s_k$ is a set of qubit pairs $\{(u,v)\}$ to be executed as simultaneous 2-qubit gates, the compilation should determine:
\begin{itemize}[leftmargin=*]
    \item Qubit placements $\pi_q^t \in \{0, 1, \ldots, M-1\}$: the trap-site index of qubit $q$ at time step $t$. Each trap index can also be decomposed as $\pi_q^t = 2\,\sigma_q^t + \hat{\ell}_q^t$, where $\sigma_q^t = \lfloor \pi_q^t / 2 \rfloor$ is the interaction site index and $\hat{\ell}_q^t = \pi_q^t \bmod 2$ is the intra-site offset.
    \item Execution stage times $\mathbf{t} = [t_0, t_1, \ldots, t_{L-1}]$: the time step at which each gate stage executes.
\end{itemize}

Any valid rearrangement must satisfy the following constraints.
\begin{enumerate}[leftmargin=*]
    \item \textit{Injectivity}: two qubits cannot occupy the same trap at any time step:
    \begin{equation*}
    \forall\, t,\; \forall\, q_1 \neq q_2 \in \mathcal{Q}: \quad \pi_{q_1}^t \neq \pi_{q_2}^t.
    \end{equation*}
    \item \textit{Scheduling precedence}: gate stages must execute in the prescribed order of the given schedule $\mathbf{S}$:
    \begin{equation*}
    \forall\, k \in \{0, \ldots, L-2\}: \quad t_k < t_{k+1}.
    \end{equation*}
    \item \textit{Gate execution}: paired qubits must share an interaction site at the scheduled time for entangling gates:
    \begin{equation*}
    % \forall\, (u, v) \in s_k: \quad \sigma_u^{t_k} = \sigma_v^{t_k}.
    \forall\, t,\forall\, k,\forall\, (u, v) \in s_k: \quad (t_k = t) \implies (\sigma_u^{t} = \sigma_v^{t}).
    \end{equation*}
    \item \textit{Site exclusivity}: idle qubits must occupy distinct interaction sites at gate stages to prevent unintended interaction, with $\mathcal{I}_k = \mathcal{Q} \setminus \bigcup_{(u,v) \in s_k} \{u, v\}$ as the set of idle qubits at stage $s_k$:
    \begin{equation*}
    % \forall\, k,\; \forall\, i \neq j \in \mathcal{I}_k: \quad \sigma_i^{t_k} \neq \sigma_j^{t_k}.
    \forall\, t,\forall\, k,\; \forall\, i \neq j \in \mathcal{I}_k: \quad (t_k = t) \implies  (\sigma_i^{t} \neq \sigma_j^{t}).
    \end{equation*}
    \item \textit{Ordering preservation}: co-moving qubits cannot cross each other during parallel transport, with $m_q^t = \mathbf{1}[\pi_q^t \neq \pi_q^{t+1}]$ denoting the movement of $q$ from $t$ to $t+1$:
    \begin{equation*}
    \forall\, t,\; \forall\, q_i, q_j:
    \left(m_{q_i}^t \wedge m_{q_j}^t\right) \implies
    \left(\pi_{q_i}^t > \pi_{q_j}^t \iff \pi_{q_i}^{t+1} > \pi_{q_j}^{t+1}\right).
    \end{equation*}
\end{enumerate}

We encode the full constraint sets as a QF\_BV satisfiability problem
to exploit the optimized bit-blasting decision procedures in modern SMT solvers. 

\subsubsection{Bidirectional Deepening}

Rather than solving a complex one-shot problem, we perform deepening over the depth parameter $T$, exploiting the monotonic structure of the problem: if no valid rearrangement exists at depth $T$, then no solution exists for any $T' < T$. We perform a bidirectional search, starting with a quick upward search to find a feasible solution, and then an iterative downward search to obtain the optimal depth. 

Specifically, there is a lower bound $T_{\text{lb}} = L$ for $L$ stages, since each gate stage requires at least one dedicated time step by scheduling constraint. In practice, we start from $T = T_{\text{lb}} + c$ for some small $c$ (e.g., $c=2$) and keep incrementing the depth with a quick timeout until delivering a feasible solution. 
Afterwards, we deepen downwards with a greater time budget to find either a better solution or to confirm optimality. This deepening will terminate when the solver returns unsatisfiable, or reaches the theoretical lower bound $T_{\text{lb}}$. In any case, a depth-optimal placement $\mathbf{P}^{(1)}$ and stage times $\mathbf{t}^{(1)}$ with $T^*$ time steps will be obtained at the end of this phase.

\subsection{Movement Compaction via MILP} \label{sec:compact}

\begin{figure*}[t]
\centering
\includegraphics[width=0.8\linewidth]{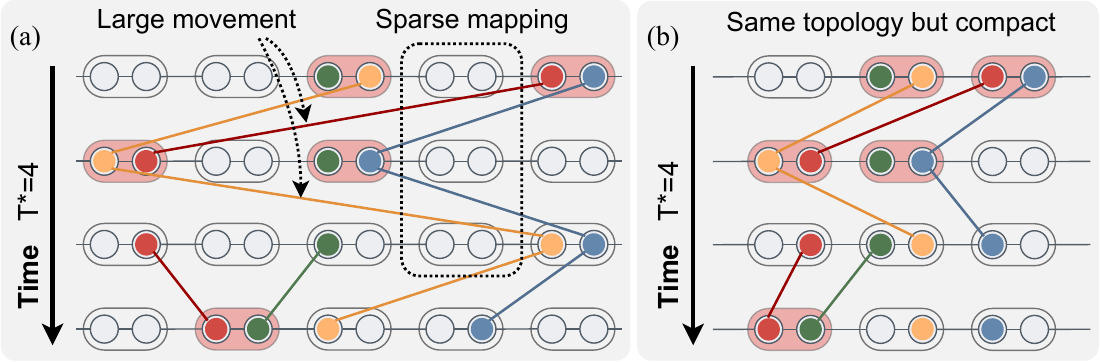}
\caption{Example solution evolution in the feedforward manner of \Name. (a) A depth-optimal combinatorial topology. (b) The physical displacement compaction.}
\label{fig:compaction}
\Description{}
\end{figure*}

The depth-optimal solution from phase 1 may not be duration-optimal due to unnecessary long movements, as shown in Figure~\ref{fig:compaction}a.
In this phase, \Name takes the depth-optimal solution as a starting point and re-assigns trap-site positions to minimize physical displacement while preserving the combinatorial topology to reduce the search space and enable efficient solving (Figure~\ref{fig:compaction}b). Since duration is dominated by the longest displacement, this compaction naturally forms a minimax problem suitable for MILP solvers.
% , leading us to adopt an MILP formulation for its exceptional ability to minimize the maximum system variables.

\subsubsection{Invariant Topology Extraction}
There are two key structural invariants from $\mathbf{P}^{(1)}$ that the MILP must preserve.
% , as highlighted and keep fixed during the compaction in Figure~\ref{fig:compaction}.

First, the permutation order $\rho^t$, which denotes the sequence of qubit indices sorted by Phase 1 placement for each time step $t$. It implicitly preserves injectivity and the no-crossing property. 

Second, the stationary pattern $\mathcal{S}^t = \{q \mid \pi_q^t = \pi_q^{t+1}\}$, which indicates for each transition $t \to t+1$, the set of qubits that do not move. This pattern prevents introducing unnecessary movements that potentially break the current compatibility.

\subsubsection{MILP Formulation}
We formulate the compaction as a MILP over the placement variables $\pi_q^t$ (integer), $\sigma_q^t$ (integer), and $\hat{\ell}_q^t$ (binary).
To encode the maximum physical displacement, 
we introduce a continuous variable $D_t$ for each transition $t \to t{+}1$.

In addition to the basic validity constraints (e.g., injectivity, gate execution, etc.) similar to SMT encoding, the general no-crossing constraint is replaced with more specific permutation order preservation and stationarity constraints to maintain the previously mentioned invariant topology.

\begin{enumerate}[leftmargin=*]
    \item \textit{Permutation order preservation}: For consecutive qubits $\rho^t(i)$ and $\rho^t(i+1)$ in the sorted order at time $t$:
    \begin{equation*}
    \pi_{\rho^t(i)}^t - \pi_{\rho^t(i+1)}^t \leq -1.
    \end{equation*}
    This constraint
    % single linear inequality per adjacent pair 
    enforces injectivity and no-crossing. 
    % simultaneously.
    \item \textit{Stationarity}: stationary qubits must remain at the same trap across the transition:
    \begin{equation*}
    \forall\, q \in \mathcal{S}^t: \quad \pi_q^t = \pi_q^{t+1}.
    \end{equation*}
\end{enumerate}

% The specific topology-preserving constraints not only ensure the solution validity in the concise integer encoding format, but also significantly reduce the search space and enable efficient solving.

The compaction objective is the movement distance. 
Due to the non-uniform trap spacing, movement distance is not simply proportional to trap-index difference. 
% To enable distance modeling within the MILP, we decompose each trap index into a site index and an intra-slot offset:
% \begin{equation}
% \pi_q^t = 2\,\sigma_q^t + \hat{\ell}_q^t, \quad \forall\, t, q.
% \end{equation}
The physical displacement of qubit $q$ during one transition is a linear function of the decomposed components, where $d_{\text{site}}$ and $d_{\text{trap}}$ denote the physical distances between adjacent sites and between the two traps within one site, respectively. We bound the maximum displacement for each transition as:
\begin{equation*}
D_t \geq \left|d_{\text{site}}\!\left(\sigma_q^t - \sigma_q^{t+1}\right) + d_{\text{trap}}\!\left(\hat{\ell}_q^t - \hat{\ell}_q^{t+1}\right)\right|, \quad \forall\, q \in \mathcal{Q},
\end{equation*}
linearized as two standard inequalities. 

This MILP carries a comprehensive compaction: first minimize the sum of maximum displacement across the rearrangement, and then minimize the total displacement as a secondary objective:
\begin{equation*}
O_1 = \min \sum_{t=0}^{T-2} D_t, \quad O_2 = \min \sum_{t=0}^{T-2} \sum_{q=0}^{N-1} a_q^t,
\end{equation*}
where $a_q^t$ introduces per-qubit displacement as auxiliary variables under the cap of $D_t$ derived from the first objective.

Note that the MILP minimizes the linear displacement rather than the actual duration computed with the kinematic model \cite{zac}:
\begin{equation}
\tau_t = 2 t_{\text{transfer}} + \sqrt{D_t/\alpha}, \label{eq:kinematic}
\end{equation}
where $t_{\text{transfer}}$ is the fixed time for trap-to-trap transfer, and $\alpha$ is the acceleration parameter. 
% The total duration is then $\sum_{t=0}^{T-2} \tau_t$.
Thus, we update the solution only if compaction improves the duration. 
% evaluate the physical duration of both the input and output and retain the original if compaction causes regression.

\begin{figure*}[t]
\centering
\includegraphics[width=0.8\linewidth]{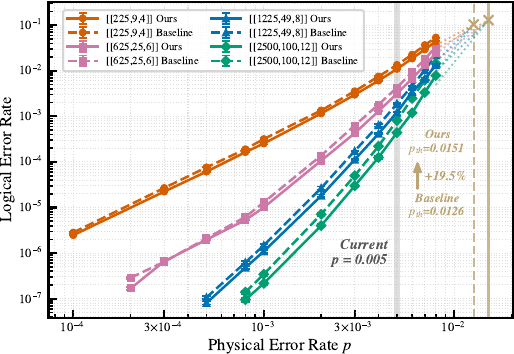}
\caption{Logical error rate (single round) versus physical error rate $p$ for four HGP codes. Solid lines denote \Name; dashed lines denote the baseline~\cite{xu_constant-overhead_2024}. Vertical lines indicate reference physical error rates (beige: estimated thresholds, gray: current representative value).
}
\label{fig:exp_lfr}
\Description{}
\end{figure*}

\subsection{Iterative Refinement via Feedback} \label{sec:feedback}

During the feedforward execution of the first two phases, a compact solution is obtained within the given depth-optimal structure. However, among the potentially vast space of depth-optimal solutions, some combinatorial topologies are inherently more amenable to compaction than others. Phase 3 therefore bridges this gap through iterative feedback between SMT and MILP, alternately tightening physical bounds and re-compacting until convergence.

% While the first two phases produce a depth-optimal solution, they do not explore the space of depth-optimal solutions with respect to rearrangement time. Phase 3 addresses this limitation by searching within the space of depth-optimal solutions to minimize rearrangement time. This is achieved through an iterative feedback loop between SMT and MILP, which progressively tightens physical constraints and re-optimizes the schedule until convergence.

\subsubsection{Local Duration Tightening}

This strategy iteratively identifies the bottleneck transition in the rearrangement and tightens its displacement bound. 
By iteratively reducing the maximum displacement, the search is guided towards more compact and duration-efficient solutions. Specifically, the displacement bound is encoded with a cap $\Delta_t^{\max}$ for a target transition $t$:
\begin{equation*}
\forall\, q \in \mathcal{Q}: \quad |\pi_q^{t+1} - \pi_q^t| \leq \Delta_t^{\max}.
\end{equation*}

\Name constructs the local search on top of the base solver from Phase 1 with all structural constraints once, and then uses incremental solving to add different displacement constraints across iterations. This allows the solver to retain learned clauses and internal heuristic state, avoiding duplicated work. 

% \input{codes/local}

% The procedure of monotonic local search is summarized in Alg.~\ref{alg:local}. At each iteration, the algorithm selects the unlocked transition with the largest displacement (line 4) and proposes a reduced cap using an adaptive step function based on its current value (line 5). To avoid re-exploring provably infeasible bounds, each transition maintains a floor, which denotes the tightest bound that has been proven infeasible so far. The tightened bounds are then asserted incrementally into the solver (line 7). 
% On SAT, both the new better solution and its displacement profile are updated and serve as the basis for the next iteration (line 8-11). On UNSAT, this over-tightened bound will be popped from the solver, and its floor will be raised accordingly (line 13). If no finer steps can be taken for tightening, this transition will be locked for the rest of the search (line 14). The loop terminates when all transitions are locked or the time budget expires, guaranteeing monotonic convergence.

At each iteration, the monotonic local search selects the transition with the largest displacement and proposes a reduced cap using an adaptive step function based on its current value. To avoid re-exploring infeasible bounds, each transition maintains a floor, which denotes the infeasible bound so far. The tightened bounds are then asserted incrementally into the solver. 
On SAT, the new solution is recorded and serves as the basis for the next iteration. On UNSAT, this over-tightened bound constraint is popped from the solver, and its floor is raised accordingly. 
If no steps can be taken for tightening, the transition is pruned for the rest of the search. The loop terminates when all transitions are saturated or the time budget is exhausted.
% , guaranteeing monotonic convergence.

\subsubsection{Global Duration Tightening}
While the local search tightens individual transitions and potentially stalls on local minima, the global search directly targets the aggregate rearrangement duration by encoding an explicit upper bound into the SMT instance.

Direct encoding of the non-linear kinematic model \eqref{eq:kinematic} within bit-vector arithmetic is not available. To address this, \Name adopts an offline pre-compute approach to bypass the complexity of non-linear constraints. 
% First, the transfer overhead is stripped from the SMT encoding first and restored analytically afterwards, since these constant shifts do not affect the optimality but help reduce bit-width requirements. This remaining movement duration depends on the physical distance, which is discretized into a finite set given the limited number of traps and qubits. 
We pre-compute the movement duration $\tau_{\delta}$ based on the trap-index difference $\delta$ and a parity bit, which indicates whether the movement crosses an interaction site boundary and thus alters the distance calculation. These values are stored as a look-up table:
\begin{equation*}
\mathcal{L}: (\delta, \text{parity}) \to \lceil \tau_{\delta} \rceil,
\end{equation*}
where $\lceil\tau_{\delta}\rceil$ represents the discretized and quantized duration for a given index displacement and parity.

Instead of $N$ independent lookups per layer, \Name identifies the transition with the largest displacement and performs a single lookup based on this bottleneck element and its parity. By performing only one table lookup, the formulation streamlines the if-then-else (ITE) chain complexity from $\mathcal{O}(N \cdot |\mathcal{L}|)$ to $\mathcal{O}(N + |\mathcal{L}|)$ per layer.
The total encoded duration $\mathcal{T}_{\text{total}} = \sum_t \mathcal{L}_t$ is represented as a zero-extended bit-vector sum.

% \input{codes/global}

% The global feedback loop is summarized in Alg.~\ref{alg:global}. Similar to the local search, the global search also maintains an unachievable floor $U$ for the duration bound. For each iteration, an adaptive step function proposes a tighter duration bound $\tau_{\text{target}}$, which is then encoded as $\hat{\mathcal{T}}$ into the solver as an upper bound constraint (line 3-5). On SAT, the new solution is evaluated for its actual duration and updated as the new best (line 6-8). On UNSAT, the proposed bound is proven infeasible and will be popped from the solver, and the floor will be raised accordingly (line 10). Convergence is declared when the gap between the best duration and the floor narrows to within the precision $\delta_{\min}$ (line 11).

Similar to the local search, the global search maintains an unachievable floor $U$ for the duration bound. For each iteration, an adaptive step function proposes a tighter duration bound $\tau_{\text{target}}$, which is then encoded into the solver as an upper bound constraint. On SAT, the new solution is evaluated and adopted as the new best. On UNSAT, the proposed bound is proven infeasible and will be popped from the solver, and the floor will be raised accordingly. Convergence is declared when the gap between the best duration and the floor narrows to within the precision.

Note that although the global search could work individually to obtain duration optimization, it is more effective when used in conjunction with the local search, as the local search can quickly tighten the transitions and provide a better starting point.

\subsection{Compilation-Time Parallelism} \label{sec:parallel}

Solver-based approaches are often criticized for their long runtimes, especially when dealing with large and complex instances. To alleviate
this concern, \Name incorporates multi-level parallelism to significantly accelerate the compilation process, making it more practical for real-world applications.

\subsubsection{Seed portfolio parallelism.}
The solver decision procedure is highly sensitive to the solving seed, with solver heuristics and branching strategies leading to vastly different search trajectories. \Name therefore launches multiple independent solver processes in parallel with identical constraints but distinct seeds. The first decisive result wins and all others are terminated. This well-known portfolio approach helps mitigate solver variability and improve overall performance.

\subsubsection{Bound speculation parallelism.} Iterative efforts happen in both Phases 1 and 3. Rather than probing bounds sequentially, \Name speculatively launches parallel calls for multiple candidate bounds, utilizing dynamic pruning and a sliding-window scheduler to reduce overall makespan.

For depth speculation (Phase 1), \Name simultaneously probes a window of candidates, \textit{e.g.}, $\{T-1, T, T+1\}$. When any depth returns SAT, all higher depths are cancelled; when a depth returns UNSAT one below the current best, optimality is immediately certified.

For duration speculation (Phase 3), \Name launches multiple duration targets spaced adaptively between the best known duration $B$ and the unachievable floor $U$. On each $B$ or $U$ update, stale tasks (targets $\geq B$ or $\leq U$) are pruned and fresh targets are submitted. This speculation process terminates when the gap $B - U$ falls within the convergence tolerance.

% \subsubsection{Composition and trade-off.} These two parallelism strategies are within the different level of compilation and can be composed to further amplify the speedup. However, they also introduce trade-offs in terms of resource usage and potential overhead from disabling incremental construction. \Name therefore provides configurable knobs to to accommodate available resources.

\section{Fidelity Analysis}
\begin{figure*}[t]
\centering
\includegraphics[width=0.8\linewidth]{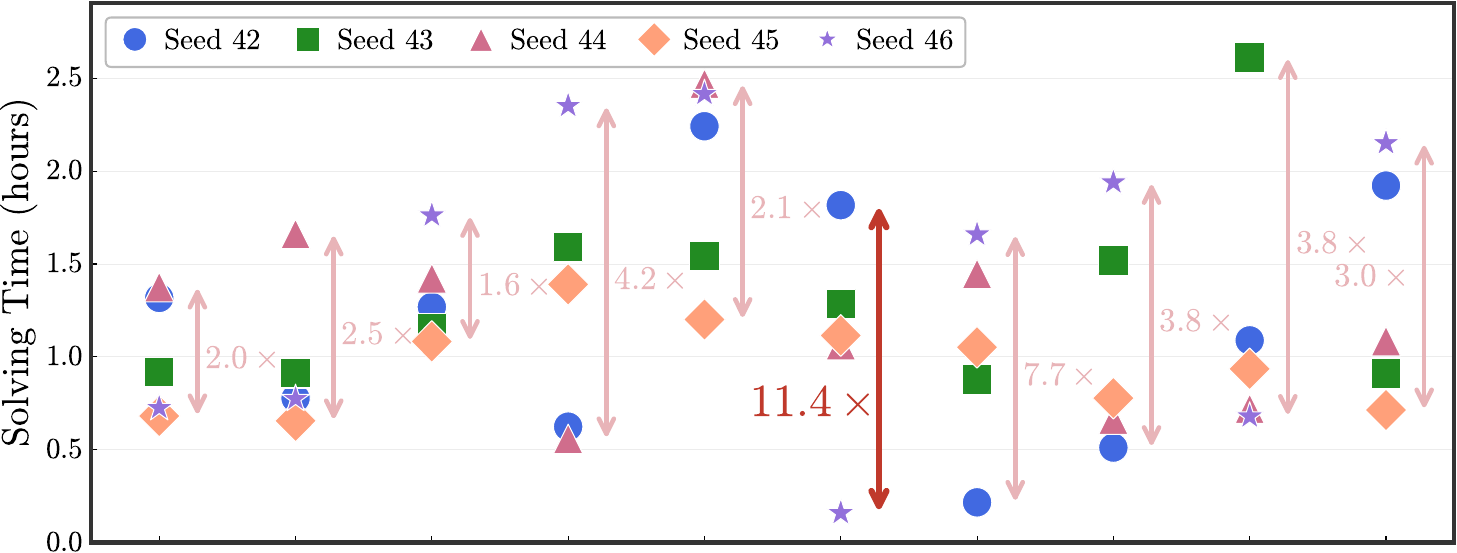}
\caption{Seed-sensitivity analysis for the SMT solver. Each dot represents the solving time for one seed on one instance; double-headed arrows annotate the max/min ratio. 
}
\label{fig:exp_portfolio}
\Description{}
\end{figure*}

Beyond direct performance gains reported in the main text, efficient physical execution in the QEC context also preserves and can even improve the logical error rate (LER) under realistic noise conditions. 
Shorter rearrangement duration directly reduces the idling error accumulated during atom transport, a factor that otherwise manifests as depolarizing noise on all qubits awaiting their next gate. 
To demonstrate this, we simulate syndrome extraction circuits for four HGP codes ($n \in \{225, 625, 1{,}225, 2{,}500\}$) under a circuit-level noise model and an additive idling error channel associated with rearrangement.
Fig.~\ref{fig:exp_lfr} plots the single-round LER versus physical error rate $p$ for both Xu et al. baseline and \Name.

By reducing the rearrangement duration per gate stage from $\sim$1.7--3\,ms to $\sim$0.3--0.8\,ms, \Name proportionally suppresses idling noise, yielding favorably consistent LER improvements across all codes and physical error rates. In the near-term fit, this more efficient execution raises the estimated threshold by 19.5\%, from $p=1.26\times10^{-2}$ to $p=1.51\times10^{-2}$. The advantage also persists in the future low-error regime, down to LERs below $10^{-7}$. At the representative physical error rate $p=0.005$, \Name reduces LER by 1.23$\times$--1.85$\times$ relative to the baseline.

The improvement is most pronounced for larger codes operating with more complex syndrome extraction, where idling noise accounts for a larger fraction of the total error budget. For the HGP~$\llbracket 2500, 100, 12 \rrbracket$ code, \Name achieves a median 43.7\% LER reduction across the swept physical error rates, while the smaller $\llbracket 225, 9, 4 \rrbracket$ code shows a more modest 17.9\% improvement. This trend underscores a critical architectural implication: as QEC codes scale to thousands of data qubits, the rearrangement overhead exerts a growing influence on the error budget. 
% Given that current simulations often neglect atom loss during transfer, the practical impact may be even larger.

Note that LER remains fundamentally constrained by the code properties. Therefore, we view the consequent LER improvement as an additional benefit beyond \Name's primary gain in execution frequency. Decoder behavior is also observed to affect the realized improvement, as local non-monotonic effects may diminish the apparent gain from reduced idling noise. We leave this interaction as an interesting topic for decoder-focused research.

\section{Parallelism Analysis}
Beyond fundamental performance metrics, we examine the specific impact of our parallelism strategies on the compilation characteristics of \Name.

\subsection{Parallelism Configuration.}
% For seed portfolio, each SMT invocation races $K_{\text{port}}$ independent solver processes: $K_{\text{port}}=10$ for Phase~1 and Phase~3 local optimization, and $K_{\text{port}}=3$ for Phase~3 global optimization. Bound speculation probes $K_{\text{spec}}$ candidate depth or duration targets in parallel, with $K_{\text{spec}}=2$ for Phase~1 and $K_{\text{spec}}=5$ for Phase~3 global optimization. These two levels of parallelism compose hierarchically, resulting in a single benchmark invocation with up to $K_{\text{port}} \times K_{\text{spec}}$ concurrent solver processes.

Table~\ref{tab:exp_config} summarizes the key parallelism parameters in the compilation experiments.
For seed portfolio, each SMT invocation races $K_{\text{port}}$ independent processes.
For bound speculation, $K_{\text{spec}}$ candidate depth or duration targets are probed simultaneously.
These two levels compose hierarchically, so a single benchmark invocation can utilize up to $K_{\text{port}} \times K_{\text{spec}}$ concurrent solver processes internally.

% Deprecated table source; retained for reference.
\begin{table}[t]
\centering
% \caption{Compilation parallelism configuration.}
\caption{Parallelism configuration for depth (Phase~1), local (Phase~3.1), and global (Phase~3.2) optimization.}
\label{tab:exp_config}
% \small
\resizebox{0.43\textwidth}{!}{%
\setlength{\tabcolsep}{2pt}
\begin{tabular}{lcccc}
\toprule
\textbf{Parameter} & {\small \textbf{Symbol}} & {\small \textbf{Depth}} & {\small \textbf{Local}} & {\small \textbf{Global}} \\ \midrule
{\small Portfolio seeds per call } & $K_{\text{port}}$ & 10 & 10 & 3 \\
{\small Speculation window size } & $K_{\text{spec}}$ & 2 & -- & 5 \\ \midrule
{\small Processes per benchmark } & -- & 20 & 10 & 15 \\
{\small Max. RAM Allocation (GB) } & -- & 100 & 50 & 75 \\
\bottomrule
\end{tabular}
}
\end{table}

\subsection{Portfolio Parallelism.}
SMT solver runtime is notoriously sensitive to the random seed used for internal branching heuristics. Fig.~\ref{fig:exp_portfolio} quantifies this variance by running five seeds on each of 10 instances at problem size $N{=}56$.

At $N{=}56$, the maximum-to-minimum runtime ratio across seeds reaches 11.4$\times$, with an average ratio of 4.2$\times$. The absolute runtime discrepancy can be as high as 1.9 hours, a gap that widens further as the scale increases. Additionally, there is no consistent ranking of seed performance across instances, indicating that heuristic performance is not intrinsically consistent but is highly sensitive to the specific underlying problem structure.

This high variance confirms that portfolio parallelism is an effective strategy for mitigating worst-case solver behavior. Racing multiple seeds and taking the first decisive solution pushes the expected wall-clock time toward the lower end of the distribution, potentially yielding an order-of-magnitude speedup and saving hours over a single-seed approach.

\begin{figure*}[t]
\centering
\includegraphics[width=0.8\linewidth]{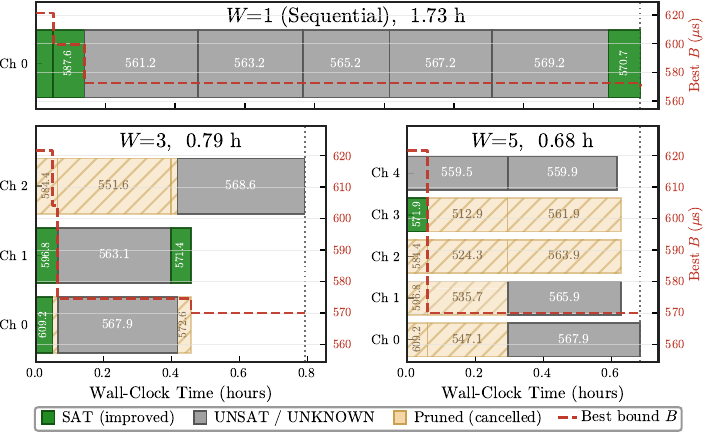}
\caption{Speculation scheduling profile for a representative instance. 
Block color denotes solver outcome (green: sat, gray: unsat/unknown, amber hatched: pruned). The dashed red line tracks the best-known duration bound $B$.
}
\label{fig:exp_spec_schedule}
\Description{}
\end{figure*}

\subsection{Speculation Parallelism.}
The iterative global search requires sequential probing of candidate duration bounds. Bound speculation parallelism accelerates convergence by launching multiple solver calls at different target bounds simultaneously.

Experimental results show that speculation achieves an average speedup of $1.9\times$ from window size $W{=}1$ to $W{=}5$, with individual cases reaching up to $3.0\times$.  
Fig.~\ref{fig:exp_spec_schedule} provides a Gantt-style visualization of the speculation scheduling for a representative instance, showing the three swim-lanes on a shared time axis. The sequential baseline ($W{=}1$) exhibits a strictly serial pattern to probe each candidate, while wider windows overlap solver tasks on parallel channels, allowing early SAT results to prune stale tasks and refill with tighter targets. 
The overall descent of the best-known bound $B$ remains consistent in all three configurations, but the wider windows reach convergence sooner by eliminating sequential bottlenecks.

%%%%%%% -- PAPER CONTENT ENDS -- %%%%%%%%

\clearpage
% \newpage
\bibliographystyle{ACM-Reference-Format}
\bibliography{ref}

\end{document}